\documentclass[bibyear]{aa}
\usepackage{graphicx}
\usepackage{xcolor}
\usepackage[normalem]{ulem}
\pdfoutput=1

\def\UC{UC~H{\sc ii}}

\def\kms{\mbox{km~s$^{-1}$}}

\def\g31{G31}
\def\stokes{Stokes\,$I$, $Q$, and $U$} 

\newcommand{\be}{\begin{equation}}
\newcommand{\ee}{\end{equation}}

\begin{document}
\title{ALMA resolves the hourglass magnetic field in G31.41+0.31} 
\author{M.\ T.\ Beltr\'an\inst{1}, M.\ Padovani\inst{1}, J.\ M.\ Girart\inst{2, 3}, D.\ Galli\inst{1}, R.\ Cesaroni\inst{1}, R.\ Paladino\inst{4}, G.\ Anglada\inst{5}, R.\ Estalella\inst{6}, M.\ Osorio\inst{5}, R.\ Rao\inst{7}, \'A.\ S\'anchez-Monge\inst{8}, Q.\ Zhang\inst{9}
}
\institute{
INAF-Osservatorio Astrofisico di Arcetri, Largo E.\ Fermi 5,
I-50125 Firenze, Italy
\and
Institut de Ci\`encies de l'Espai (ICE, CSIC), Can Magrans s/n, E-08193 Cerdanyola del Vall\`es, Catalonia, Spain
\and
Institut d'Estudis Espacials de de Catalunya (IEEC), E-08034 Barcelona, Catalonia, Spain
\and
INAF-Istituto di Radioastronomia, via P.\ Gobetti 101, I-40129 Bologna, Italy 
\and
Instituto de Astrof\'{\i}sica de Andaluc\'{\i}a, CSIC, Glorieta
de la Astronom\'{\i}a, s/n, E-18008, Granada, Spain
\and
Departament de F\'{\i}sica Qu\`{a}ntica i Astrof\'{\i}sica, Institut de Ci\`{e}ncies del Cosmos, Universitat de Barcelona, IEEC-UB, Mart\'{\i} i Franqu\`{e}s 1, E-08028 Barcelona, Spain
\and
Institute of Astronomy and Astrophysics, Academia Sinica
645 North Aohoku Place, Hilo, Hawaii 96720, USA
\and
I.\ Physikalisches Institut, Universit\"at zu K\"oln, Z\"ulpicher Str.\ 77, D-50937 K\"oln, Germany
\and
{Center for Astrophysics \textbar\ Harvard \& Smithsonian, 60 Garden Street, Cambridge, MA 02138, USA}
}
\offprints{M.\ T.\ Beltr\'an, \email{mbeltran@arcetri.astro.it}}
\date{Received date; accepted date}

\titlerunning{Polarization in G31.41+0.31}
\authorrunning{Beltr\'an et al.}

\abstract
{Submillimeter Array (SMA) 870\,$\mu$m polarization observations of the hot molecular core G31.41+0.31 revealed one of the clearest examples up to date of an hourglass-shaped magnetic field morphology in a high-mass star-forming region.}
{To better establish the role that the magnetic field plays in the collapse of G31.41+0.31, we carried out Atacama Large Millimeter/submillimeter Array (ALMA) observations of the polarized dust continuum emission at 1.3\,mm with an angular resolution four times higher than that of the previous (sub)millimeter observations to achieve an unprecedented image of the magnetic field morphology.}
{We used ALMA to perform full polarization observations at 233\,GHz (Band 6). The resulting synthesized beam is $0\farcs28\times0\farcs20$ which, at the distance of the source, corresponds to a spatial resolution of $\sim$875\,au.}
{The observations resolve the structure of the magnetic field in G31.41+0.31 and allow us to study the field in detail.  The polarized emission in the Main core of G31.41+0.41is successfully fit with a semi-analytical magnetostatic model of a toroid supported by magnetic fields. The best fit model suggests that the magnetic field is well represented by a poloidal field with a possible contribution of a toroidal component of $\sim$10\% of the poloidal component, oriented southeast to northwest at $\sim -44^\circ$ and with an inclination of $\sim-45^\circ$. The magnetic field is oriented perpendicular to the northeast to southwest velocity gradient detected in this core on scales from 10$^3$--10$^4$\,au. This supports the hypothesis that the velocity gradient is due to rotation of the core and suggests that such a rotation has little effect on the magnetic field. The strength of the magnetic field estimated in the central region of the core with the Davis-Chandrasekhar-Fermi method is $\sim$8--13\,mG and implies that the mass-to-flux ratio in this region is slightly supercritical ($\lambda$=1.4--2.2).}
{The magnetic field in G31.41+0.31 maintains an hourglass-shaped morphology down to scales of  $<$1000\,au. Despite the magnetic field being important in G31.41+0.31, it is not enough to prevent fragmentation and collapse of the core, as demonstrated by the presence of at least four sources embedded in the center of the core.}

\maketitle

\section{Introduction}

The idea that magnetic fields play a dynamically important role in the process of star formation has been advocated for many years (e.g., Shu et al.~\cite{shu99}; Mouschovias \& Ciolek~\cite{mouschovias99}). However, in recent years, the validity of magnetically dominated star formation theories and models has been questioned by theories arguing that the star formation process is driven by turbulent flows, especially in the high-mass regime (e.g., Mac Low \& Klessen~\cite{maclow04}). This leads to a situation in which, despite decades of research, no consensus regarding the importance of magnetic fields in star formation has been reached. The advent of polarization observations has started to change this situation because they have proven to be an excellent tool to measure the direction of the magnetic field in star-forming regions (e.g., Li et al.~\cite{li09}; Davidson et al.~\cite{davidson11}; Hull et al.~\cite{hull13}; Zhang et al.~\cite{zhang14}; Hull \& Zhang~\cite{hull19}) and to assess the relative magnitudes of the mean and turbulent components of the field (Hildebrand et al.~\cite{hildebrand09}).

In the low-mass regime, Girart et al.~(\cite{girart06}) carried out sub-millimeter polarization observations toward NGC\,1333 IRAS\,4A and reported a {\it textbook case} of an hourglass-shaped magnetic field morphology. This is expected in the supercritical regime of the core collapse when an initially uniform magnetic field is advected and compressed by the accreting material. A detailed analysis of the polarization data shows that the IRAS\,4A magnetic field morphology is consistent with the prediction in the standard core collapse models for magnetized clouds (Galli \& Shu~\cite{galli93a},\cite{galli93b}; Fiedler \& Mouschovias~\cite{fiedler93}; Gon\c{c}alves et al.~\cite{goncalves08}; Frau et al.~\cite{frau11}).

In the high-mass regime, one of the clearest examples up to date of an hourglass-shaped magnetic field morphology is that of the hot molecular core (HMC) G31.41+0.31 (hereafter G31; Girart et al.~\cite{girart09}, hereafter GIR09). G31 is a massive HMC core ($>$100\,$M_\odot$, $\sim$40--1200 K: Beltr\'an et al.~\cite{beltran04}, \cite{beltran05}, \cite{beltran18}, hereafter BEL18; GIR09; Osorio et al.~\cite{osorio09}; Cesaroni et al.~\cite{cesa11}) with a luminosity of $\sim 2\times10^5 L_\odot$ (Osorio et al.~\cite{osorio09}), located at a kinematic distance of $\sim7.9$\,kpc, and thought to be heated by one or more O-B (proto)stars. New parallax observations (Reid et al., in prep) have located this high-mass star-forming region much closer, at 3.7\,kpc. Therefore, the luminosity of the region would be of $\sim 4.4\times10^4 L_\odot$, and the core mass $\sim 26\,M_\odot$, with this new distance, according to BEL18. This estimate was obtained from very high angular resolution interferometric observations, and, therefore, it should be taken as a lower limit  because the interferometer might have filtered spatially extended emission. Taking the accuracy of the parallactic distances into account, from now on we use 3.7\,kpc as the distance to G31.  Centimeter Very Large Array (VLA) observations reveal two continuum sources close to the center of the HMC and separated by $\sim 0\farcs2$ (Cesaroni et al.~\cite{cesa10}). Line emission observations show that the core simultaneously rotates and infalls (GIR09; Mayen-Gijon et al.~\cite{mayen14}; BEL18), while $1''$ angular resolution dust polarization observations carried out at 870~$\mu$m with the Submillimeter  Array (SMA)  reveal that the magnetic field lines threading the HMC are pinched along its major axis, acquiring the characteristic hourglass shape (GIR09). These submillimeter observations also reveal that the magnetic field dominates centrifugal and turbulence forces in the dynamics of the collapse and that the rotation velocity of the core decreases for decreasing radii (GIR09). This suggests that magnetic braking may transfer angular momentum in the core (Basu \& Mouschovias~\cite{basu94}; Galli et al.~\cite{galli06}; Mellon \& Li~\cite{mellon08}). However, this latter scenario has recently been challenged by new high-angular resolution ($0\farcs2$) Atacama Large Millimeter/submillimeter Array (ALMA) observations at 1.4\,mm that suggest that the rotation in G31 spins up close to the center (BEL18).

The HMC in G31 is a factor of 20 larger and more massive, and four orders of magnitude more luminous than the Sun-like object NGC\,1333 IRAS\,4A. In spite of this, the magnetic field characteristics of the two sources are similar: an hourglass configuration (suggesting that the envelope might be partially supported by the field while contracting preferentially along B-field lines), a similar mass-to-flux ratio, and a magnetic field energy dominating over turbulence. This similarity suggests that the role of magnetic field in the early stages of the formation of high- and low-mass stars may not be too different. However, IRAS\,4A is hosting a binary protostellar system with a separation of 500 AU, and therefore, the B-field properties estimated by Girart et al.~(\cite{girart06}) and the dynamics of the collapse are directly related to those of the low-mass protostar(s). On the other hand, G31 will probably form a group of stars or small cluster, as
suggested by the mass of the core, $>26\,M_\odot$ (BEL18), and the presence of two centimeter continuum embedded sources detected by Cesaroni et al.~(\cite{cesa10}). Recent $0\farcs08$ angular resolution ALMA observations at 1.4 and 3.5\,mm (2016.1.00223 -- PI: M.\ Beltr\'an) resolve the dust continuum emission toward the center of G31 for the first time (Beltr\'an et al., in prep.). These observations clearly reveal the presence of at least four embedded sources, two of which associated with the centimeter continuum sources previously detected by Cesaroni et al. (2010).

To better establish the role that the magnetic field plays in the collapse of G31, we carried out ALMA observations of the polarized dust continuum emission at 1.3\,mm (Band 6) with an angular resolution of $\sim 0\farcs2$, which is four times higher than that of previous (sub)millimeter observations (GIR09) and is similar to the separation of the centimeter continuum sources embedded in the core (Cesaroni et al.~\cite{cesa10}). This allowed us to achieve an unprecedented image of the magnetic field morphology down to $\sim$800\,au scales.

In  this work, we analyze the ALMA observations and compare them to semi-analytical magnetostatic models of a toroid supported by magnetic fields.  In Sect.\ 2 we describe the ALMA observations; in Sect.\ 3 we present the results on the continuum and polarized emission toward G31; in Sect.\ 4
we model the magnetic field and estimate its strength using the method by Davis~(\cite{davis51}) and Chandrasekhar \& Fermi~(\cite{chandra53}); in Sect.\ 5 we discuss whether the G31 core is supercritical based on the mass-to-flux ratio, as well as the influence of rotation on the magnetic field and the possible causes for the deviation between our best model and the data. Finally, in Sect. 6 we give
our main conclusions.

\begin{figure}
\includegraphics[angle=0,width=8.5cm]{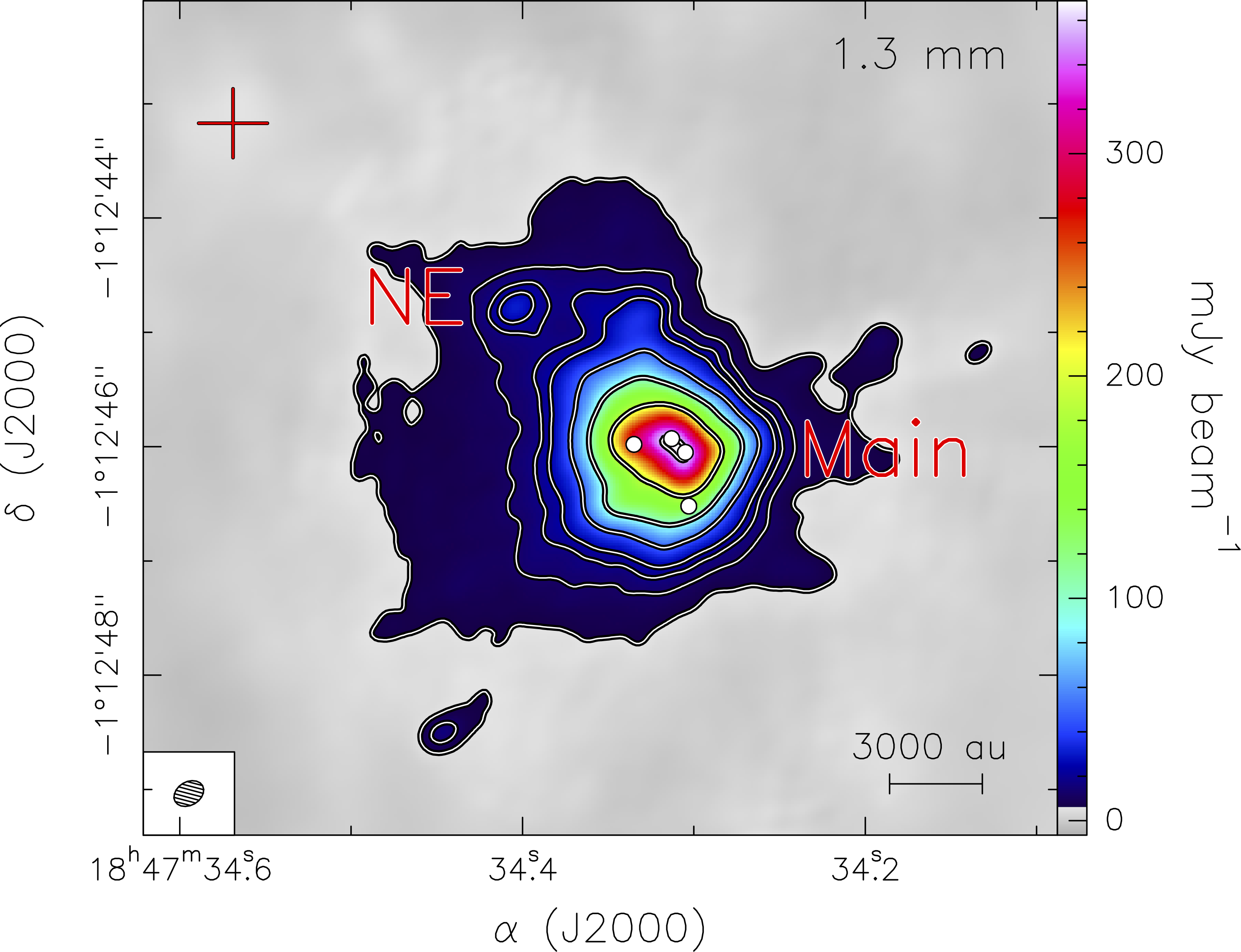}
\vspace{.2cm}
\includegraphics[angle=0,width=7.3cm]{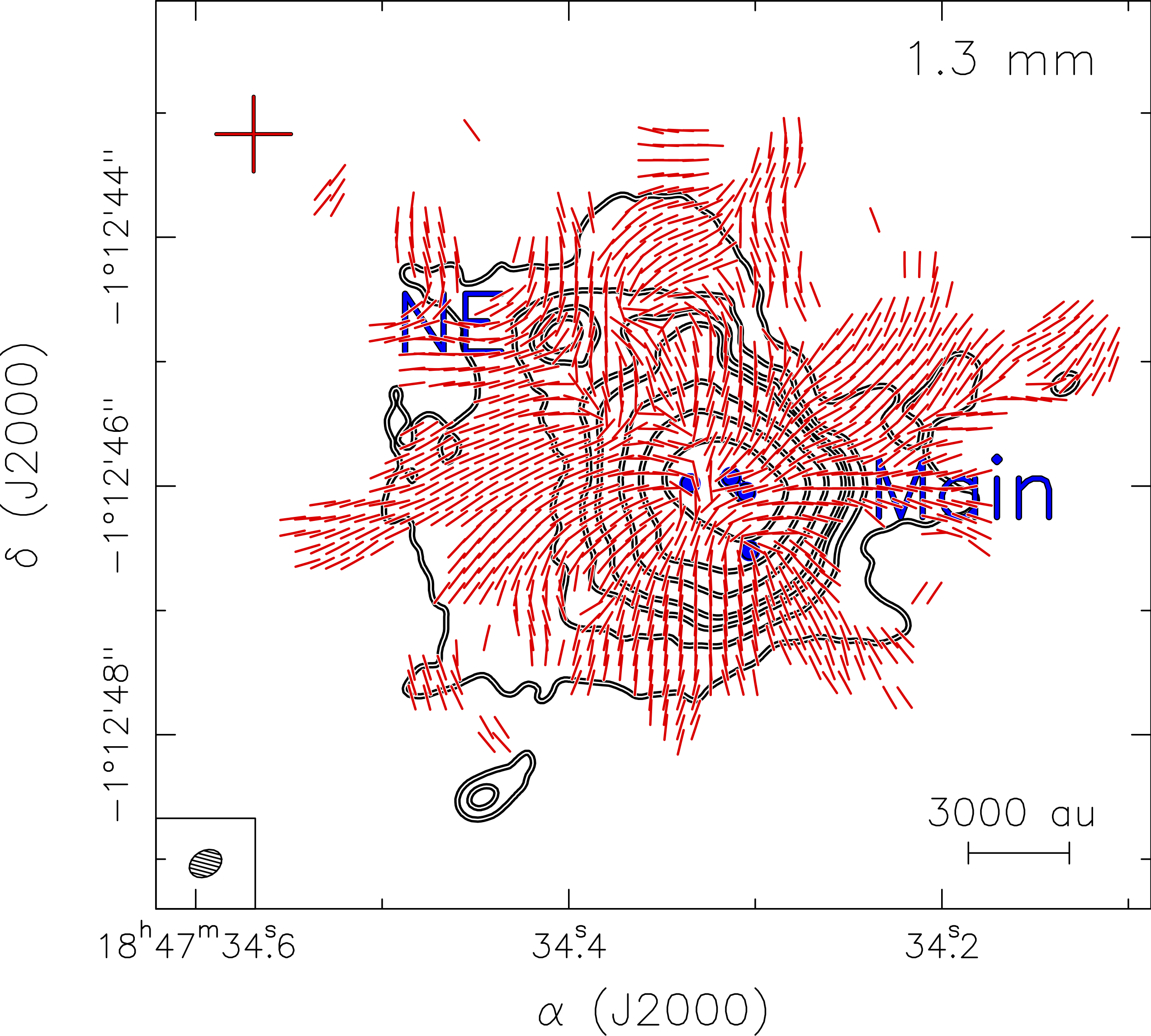}
\caption{({\it Top panel}) 1.3\,mm ALMA continuum emission map of the HMC G31.  The contours are 5, 10, 15, 20, 40, 80, 160, and 300 times $\sigma$, where 1$\sigma$ is 1.2\,mJy\,beam$^{-1}$.  The white dots mark the position of four embedded continuum sources observed at 1.4\,mm and 3.5\,mm (Beltr\'an et al., in prep). The red cross indicates the position of the \UC\ region imaged by Cesaroni et al.~(\cite{cesa94}).
The synthesized beam is shown in the lower left-hand corner. 
({\it Bottom panel})  Magnetic field segments (red lines), obtained by rotating $90^\circ$ the polarization segments, overlapped on the 1.3\,mm continuum emission map (contours). The segments are sampled following Nyquist sampling (every five pixels). }
\label{fig-cont}
\end{figure}

\section{Observations}\label{sect:obs}

Interferometric full polarization observations of \g31 were carried out with ALMA in Cycle 4 on July 12, 2017 as part of project 2015.1.00072.S (P.I.: M.\ Beltr\'an). The total observing time was divided into two execution blocks.  We used ALMA in full polarization mode and observed all four cross correlations using a spectral setup with four 2\,GHz spectral windows of 64 channels each (TDM mode with 31.25\,MHz resolution per channel). From the (XX, XY, YX, and YY) visibilities we obtained the Stokes\,$I$, $Q$, and $U$ in the image plane. The observations were performed in Band 6 centered at 233\,GHz and with the array in the C40--5 configuration.  The baselines of the observations range from $\sim$17 to 2647\,m.
 The phase reference center of the observations is $\alpha$(J2000)$=18^{\rm h}\,47^{\rm m}\,34\fs308$, $\delta$(J2000)$=-01^\circ\,12^\prime\,45\farcs90$. 
 Phase calibration was performed on quasar J1851+0035, while flux and bandpass calibrations were performed on quasar J1751+0939.  Quasar J1924$-$2914 was observed to determine the instrumental contribution to the cross-polarized interferometer response. 

The data were calibrated and imaged using the {\sc CASA}\footnote{The {\sc CASA} package is
available at \url{http://casa.nrao.edu/}} software package (McMullin et al.~\cite{mcmullin07}). Further imaging and
analysis were done with the {\sc GILDAS}\footnote{The {\sc GILDAS} package is available at
\url{http://www.iram.fr/IRAMFR/GILDAS}} software package.  
We performed self-calibration using the total intensity (Stokes\,$I$) image
as a model but the images did not improve and we decided 
not to apply it. Therefore, the data presented here have not been self-calibrated. The final maps were created using the CLEAN task with natural weighting. The resulting synthesized CLEANed beam of the maps is $0\farcs28\times0\farcs20$ at a position angle PA of $-60^\circ$. 
The rms noise of the maps is 1.2\,mJy\,beam$^{-1}$ for Stokes\,$I$ and 22\,$\mu$Jy\,beam$^{-1}$ for Stokes\,$Q$ and $U$. The fact that the rms noise of Stokes\,$I$ is a factor $\sim 50$ higher than that of Stokes\,$Q$ and $U$ is due to a problem of imaging dynamic range, because the dynamic range for Stokes\,$I$ is $>300$.

From the \stokes, we derived the linear polarization intensity, $P=\sqrt{Q^2+U^2}$, the fractional linear polarization, $p=P/I$, and the polarization position angle, $\psi=\frac{1}{2}\arctan(U/Q)$.  The accuracy of the polarization position angle  $\psi$ is of a few degrees while that of the  fractional linear polarization $p$ is $\sim 0.1$\%. Assuming that the polarization is produced by magnetically aligned dust grains, in all the figures we show polarization segments rotated by $90^\circ$ to outline the direction of the magnetic field.

\section{Results}

\subsection{Continuum emission}
\label{sect:cont}

Figure~\ref{fig-cont} shows the map of the Stokes\,$I$, namely, the total intensity at 1.3\,mm, in G31. The intensity map is consistent with the 1.4\,mm dust continuum emission map obtained by BEL18 with  similar angular resolution. The two cores detected by BEL18 and named NE and Main are clearly visible. Thanks to the higher sensitivity of our polarized observations, we could image low-intensity extended emission surrounding both cores and  additional cores, such as the one located $\sim$3$''$ to the southeast of the Main core.  The morphology of the low-intensity (5$\sigma$) emission resembles that observed with the SMA at 870\,$\mu$m and $1''$ angular resolution. As already noticed by BEL18, the Main core appears rather uniform and compact, with no hints of fragmentation despite the fact that the synthesized beam is at least ten times less than the core diameter.  However, as suggested by these authors, the homogeneous and monolithic appearance of the core is probably due to a combination of large dust continuum opacity and insufficient angular resolution to resolve the small-scale structure of the extended component. In fact, new ALMA observations at $\sim$0$\farcs08$ have resolved the core in at least four embedded sources, confirming that fragmentation has already taken place in G31 (Beltr\'an et al., in prep.).
 
The total flux measured inside the 5$\sigma$ contour level, which has a size of $\sim$4$^{\prime\prime}$ or $15000$\,au at the distance of the source, is $6.86 \pm 0.06$\,Jy, while the peak flux is $369 \pm 1$\,mJy\,beam$^{-1}$. Because of the TDM mode of the observations, which provides limited spectral resolution ($\sim$40\,\kms), and the intense line emission of G31 (e.g., Beltr\'an et al.~\cite{beltran05}; Rivilla et al.~\cite{rivilla17}), the total flux measured for the Stokes\,$I$ should be considered an upper limit. To estimate the line contamination, we used the dust continuum emission flux obtained by BEL18, who carried out observations at 217\,GHz with a similar angular resolution ($\sim$0$\farcs22$) but much higher spectral resolution (2.7\,\kms) which allowed them to properly determine the continuum level. The dust continuum emission inside the $5\sigma$ contour level at 217\,GHz is 3.75\,Jy. We estimated the continuum dust emission flux at 233\,GHz from that at 217\,GHz adopting the scaling $S_\nu\propto\nu^\alpha$, where the spectral index $\alpha=2+\beta$ and $\beta$ is the dust emissivity index. We used two different values of the dust emissivity index, $\beta=1$ and 2, and estimated a dust continuum emission flux of 4.64 and 4.98\,Jy, respectively. Therefore, the line contamination of the total flux at 233\,GHz would be of 48\% for $\beta=1$ and 38\% for $\beta=2$. We conclude that at least $\sim$40\% of the total flux estimated for Stokes\,$I$ is contaminated by line emission.

\begin{figure}
\centerline{\includegraphics[angle=0,width=8.5cm]{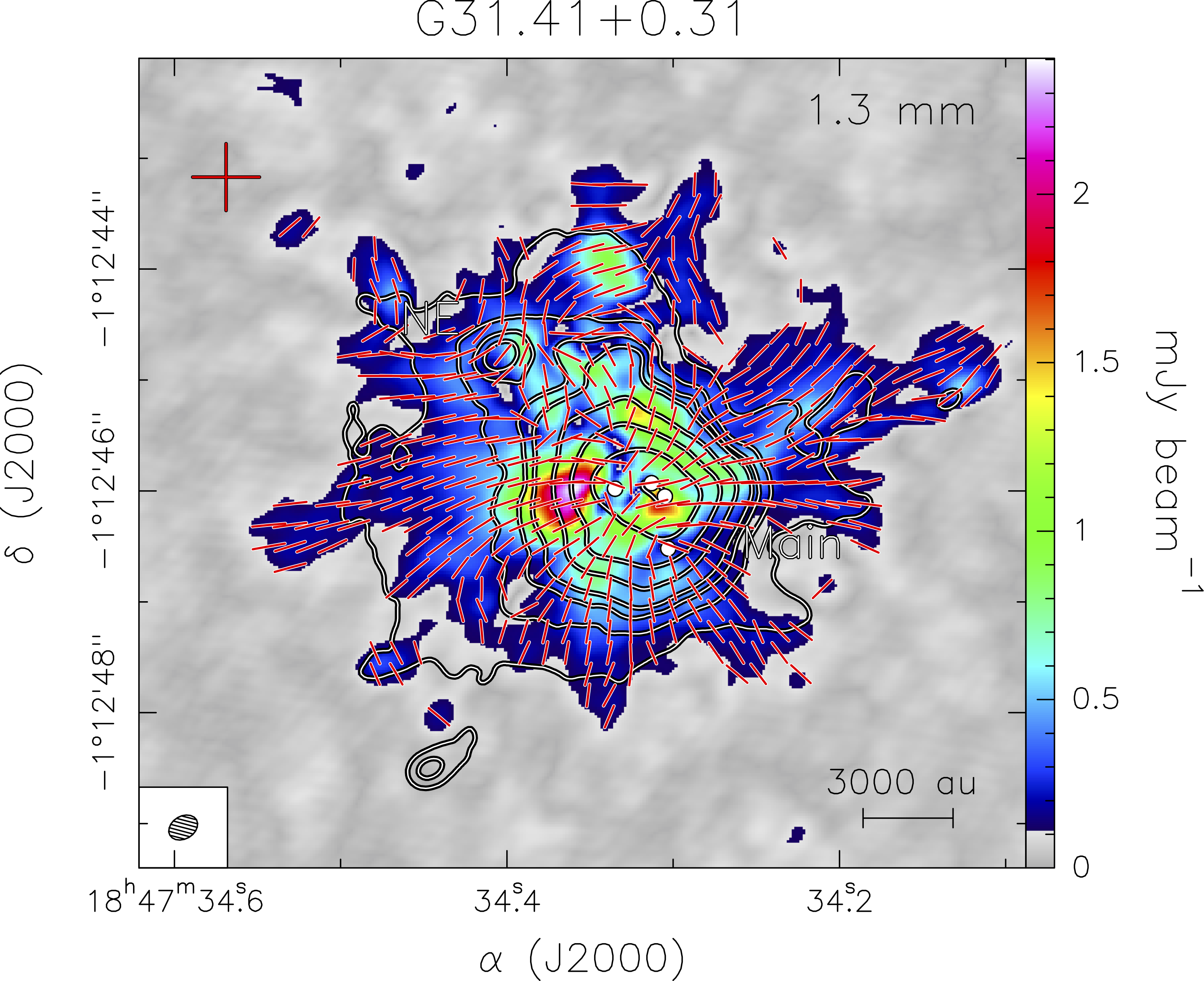}}
\hspace{-.2cm}
\centerline{\includegraphics[angle=0,width=8.4cm]{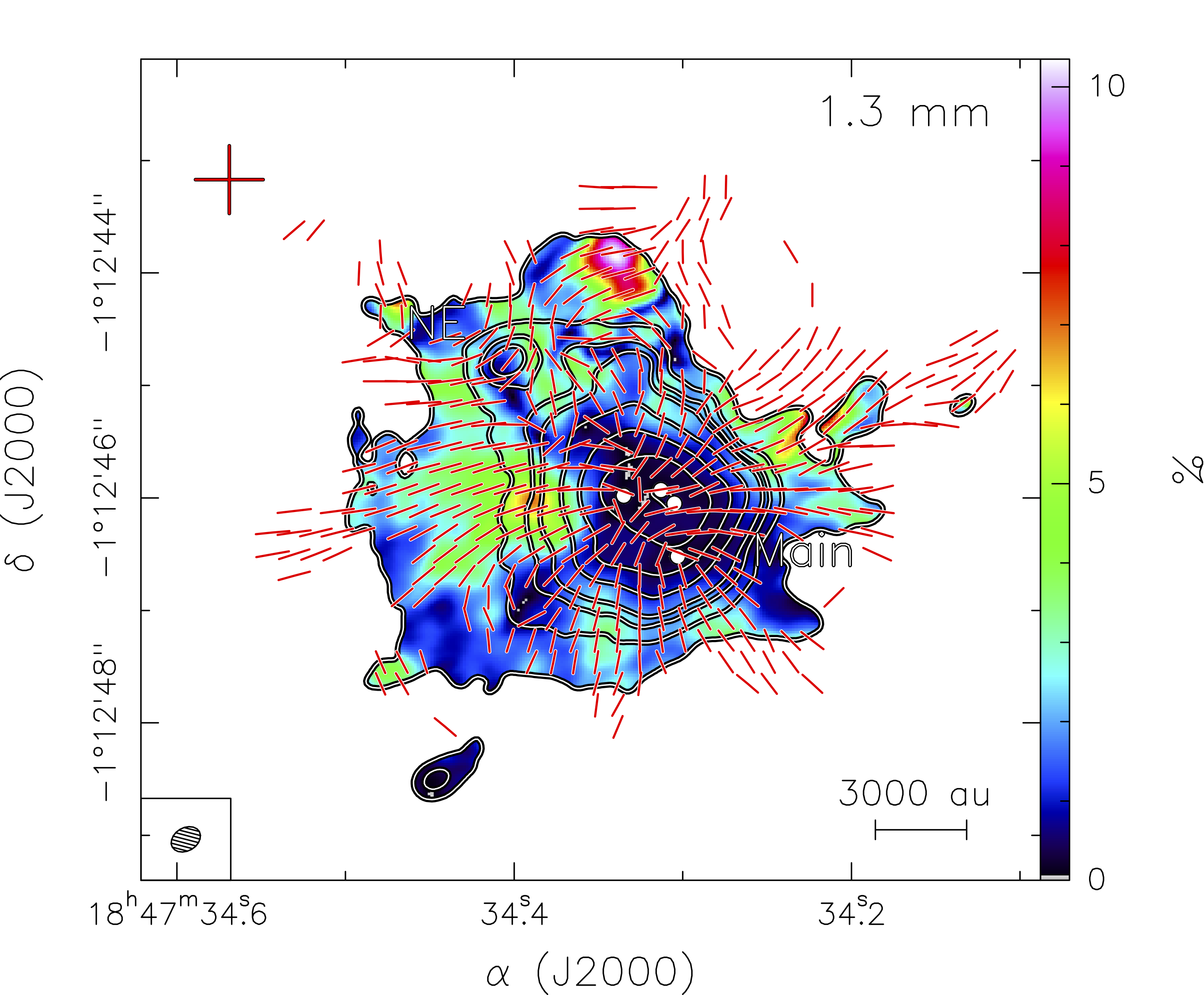}}
\caption{({\it Top panel}) Linearly polarized intensity $P$ (colors) and dust continuum emission map (contours) at 1.3\,mm in G31. Polarized intensity ranges from 0.11\,mJy\,beam$^{-1}$ to 2.4\,mJy\,beam$^{-1}$. ({\it Bottom panel}) Polarization fraction $p$ (colors) and dust continuum emission map (contours) at 1.3\,mm.  Polarization fraction ranges from 0.1\% to 13\%. The red thick segments in both panels indicate the magnetic field lines.  Segments are shown every eight pixels. The synthesized beam is shown in the lower left-hand corner.  Contours and symbols are the same as in Fig.~\ref{fig-cont}. }
\label{fig-pol}
\end{figure}

\subsection{Polarized emission}

Figure~\ref{fig-pol} shows the linearly polarized emission, $P$, and the polarization fraction, $p$, in G31. We detected linearly polarized dust emission mainly in the Main core in G31, with a maximum polarized flux of  2.4\,mJy\,beam$^{-1}$ eastward of the dust continuum emission peak and of the millimeter and centimeter continuum embedded sources. The polarization fraction in the HMC ranges  from 0.1\% to 13\%. The maximum fraction level is found outside the Main core toward the northeast and the northwest.  A secondary peak, with a polarization fraction at a $\sim$7\% level, is located eastward of the dust continuum peak. This enhancement of the fractional polarization is associated with the peak of the polarized emission.
The fractional polarization decreases to a 0.15--0.5\% level toward the position of the dust continuum emission peak and of the two centimeter continuum embedded sources. Despite the presence of a relative maximum of polarized intensity maximum near the Stokes\,$I$ maximum, the polarization fraction does not exceed 0.5\%. We note that because the total flux measured for Stokes\,$I$ should be considered as an upper limit due to line contamination (see previous section), the polarization fraction has to be taken as a lower limit. This is probably more important at the center of the core than close to the border of it.

The polarized emission could trace either magnetic fields or dust scattering (e.g., Girart et al.~\cite{girart06}; Kataoka et al.~\cite{kataoka15}). In G31, the polarization pattern at 1.3~mm is consistent with that observed at 870\,$\mu$m with the SMA at an angular resolution of $1''$.  
Moreover, in many parts of the outer envelope the polarization fraction reaches values of 4--5\%. 
This  suggests that the polarization observations are likely tracing the emission of magnetically-aligned grains and are not affected by dust scattering (e.g., Alves et al.~\cite{alves18}), because in the latter case the polarization should change with wavelength and the polarization fraction should be smaller. 
Only in the inner part of the core (radius of $\la0\farcs5$),  the polarization fraction is very low, $\sim 0.5$\%. In this case, self-scattering could be significant only if the maximum grain size is at least 50\,$\mu$m (see Kataoka et al. \cite{kataoka17}). However, the observations are sensitive to scales from $\sim 10^3$\,au to $10^4$\,au, so it is unlikely that dust grains at such scales have grown to the sizes observed in circumstellar disks at $\sim$100\,au scales or lower (tens to hundreds of $1\,\mu$m: e.g., Girart et al.~\cite{girart18}; Bacciotti et al.~\cite{bacciotti18}) for which dust scattering has been observed (Girart et al.~\cite{girart18}; Bacciotti et al.~\cite{bacciotti18};  Hull et al.~\cite{hull18}; Dent et al.~\cite{dent18}), although there are exceptions (see Alves et al.~\cite{alves18}).

\begin{figure}
\centerline{\includegraphics[angle=0,width=8.5cm]{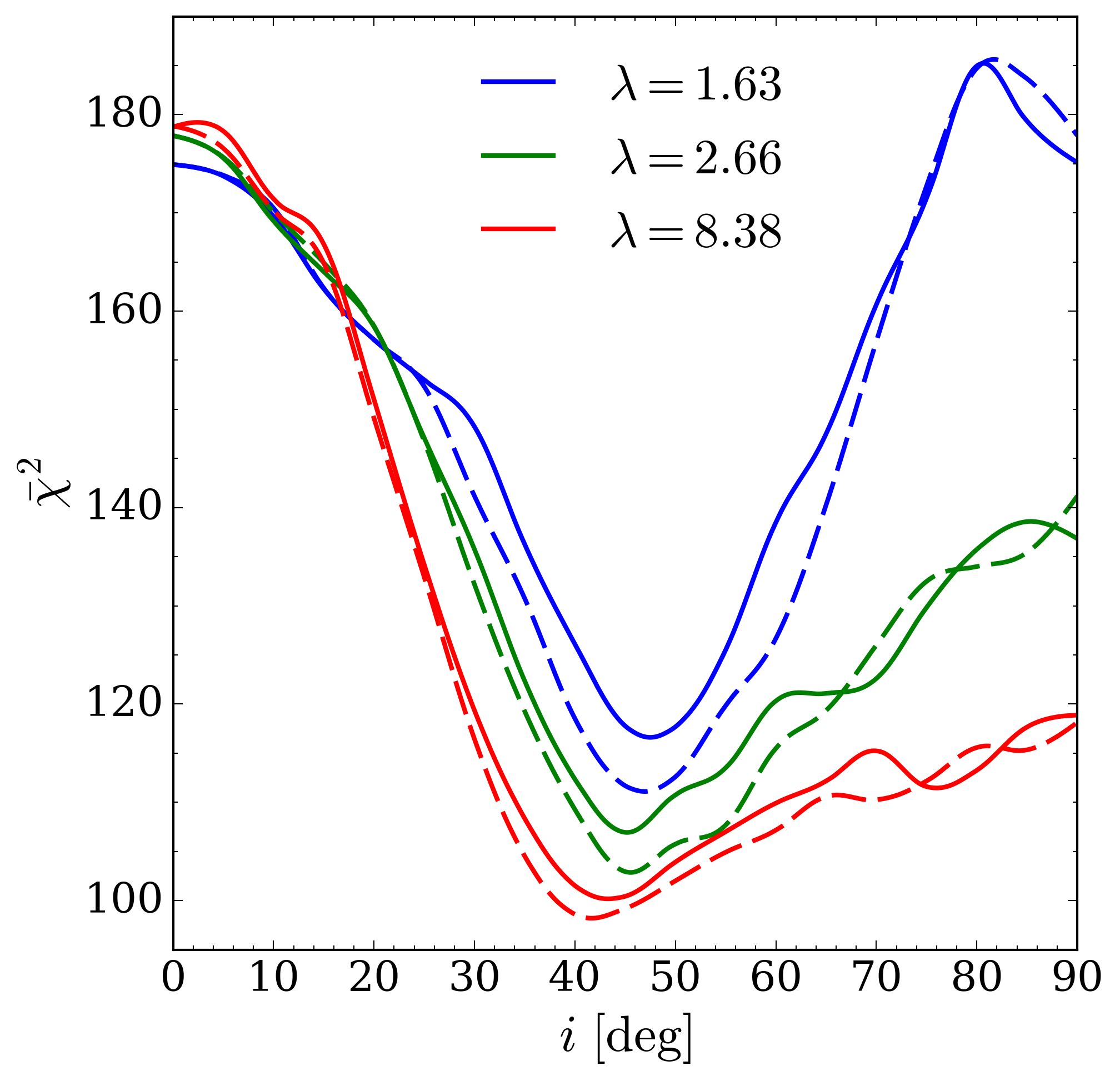}}
\hspace{-.2cm}
\centerline{\includegraphics[angle=0,width=8.4cm]{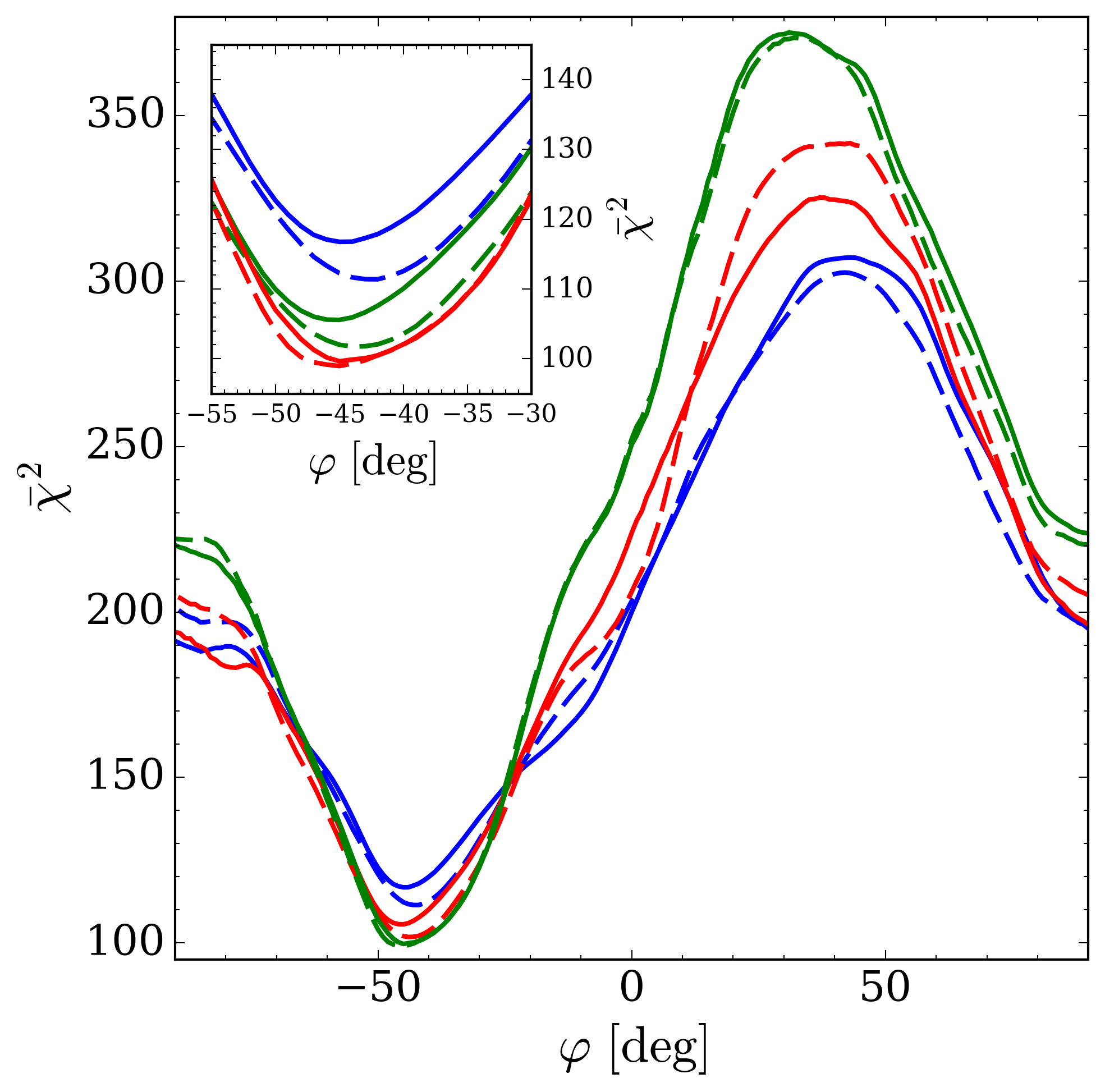}}
\caption{{\it Top panel}: Minimum reduced chi squared versus the inclination, $i$, 
of the model for three different values of the
mass-to-flux ratio, $\lambda$, and $b_0=0.1$.
Solid (dashed) lines show $\bar\chi^2$ for positive (negative) values of $i$. For a better comparison,
negative inclinations are shown in absolute value.
{\it Bottom panel}: reduced chi squared versus the orientation on the plane of the sky for $i=-45^\circ$. The inset shows a zoom around the minimum of $\bar\chi^2$.
The color coding follows that of the upper panel.}
\label{chi2vsinclphi}
\end{figure}

\begin{figure*}[!h]
\centerline{\includegraphics[angle=0,width=18cm,angle=0]{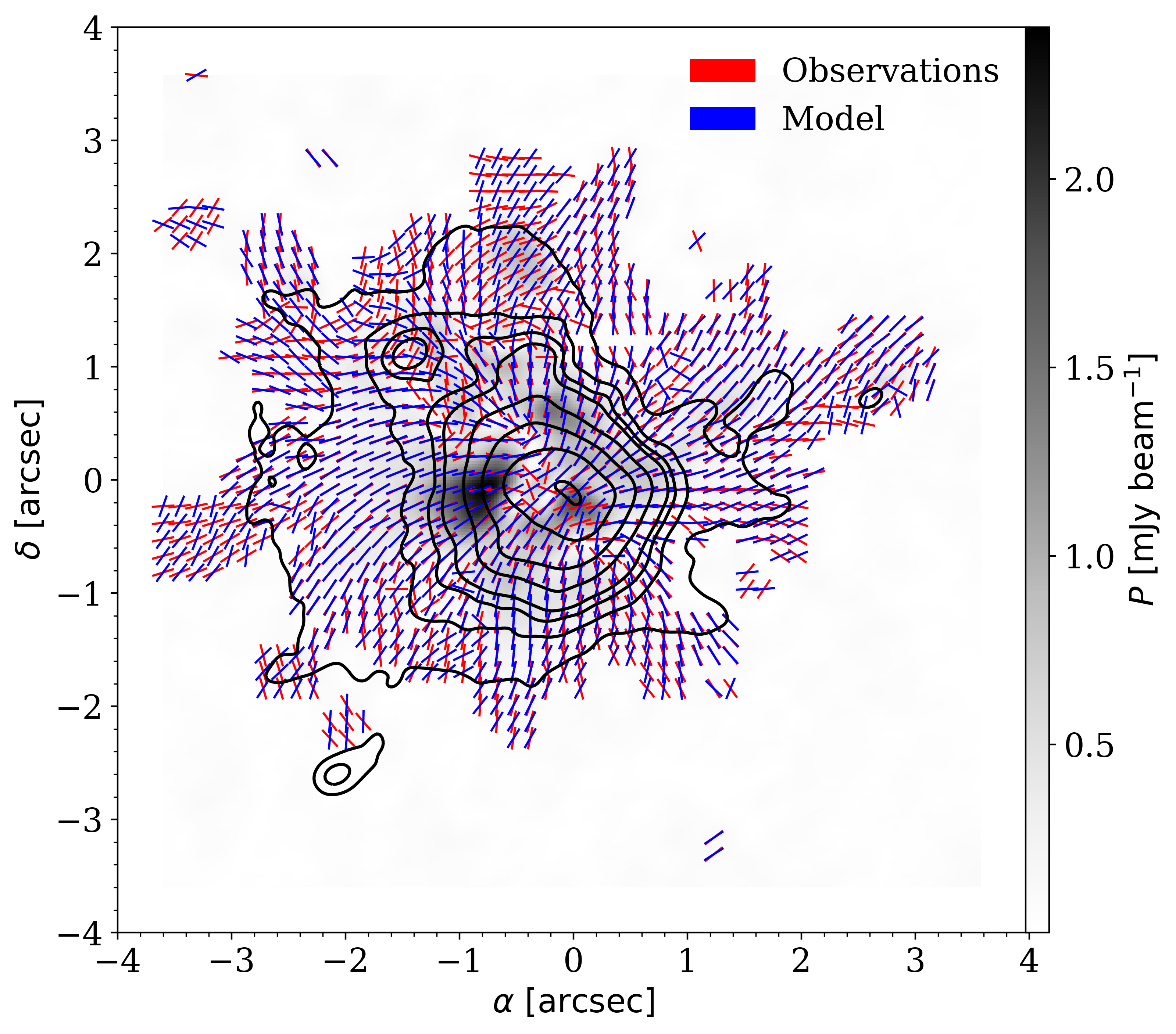}}
\caption{Polarization angles showing the magnetic field direction from observations ({\it red})
and the best model ({\it blue}: $b_0=0.1$, $\lambda$=2.66, $i=-{45^\circ}^{+3^\circ}_{-4^\circ}$, and $\varphi=-{44^\circ}_{-4^\circ}^{+6^\circ}$). The gray-scale map shows the polarized intensity, $P$, while
black contours show the 1.3~mm dust emission
at 5, 10, 15, 20, 40, 80, 160, and 300 times $\sigma$, (see Fig.~\ref{fig-cont} for details).  We note that although the magnetic field segments of the observations are shown also for regions below the 5$\sigma$ contour, these have not been used for the calculations or the discussion.}
\label{vectormap}
\end{figure*}

\section{Analysis}

\subsection{Modeling the magnetic field}
\label{sect:model}

Assuming that the polarization pattern is due to dust grains with their shortest axis 
aligned to the magnetic field,  we used the {\em DustPol} module of the ARTIST package (Padovani et al.~\cite{padovani12})
to model the magnetic field morphology. 
We modeled G31 as an axially-symmetric singular 
toroid threaded by a poloidal magnetic field (Li \& Shu~\cite{li96}; Padovani 
\& Galli~\cite{padovani11}). 
Following Padovani et al.~(\cite{padovani13}), we added a toroidal 
force-free component of the magnetic field, modified 
here to mimic the effects of rotation. In practice, we allowed the 
magnetic field to make a ``kink'' at the midplane, assuming 
opposite signs of the toroidal component above and below 
the midplane. The ratio of the toroidal and poloidal components
of the field remains approximately
constant along each field line, except close to the magnetic 
axis, which coincides with the rotation axis, where the toroidal component dominates (see Padovani et al.~\cite{padovani13} for details).

This model has four free parameters: $(i)$ the mass-to-flux ratio, $\lambda$, defined as
\be\label{m2fratio}
\lambda=2\pi G^{1/2}\frac{M(\Phi)}{\Phi}\,,
\ee
where $G$ is the gravitational constant, $\Phi$ the magnetic flux, and $M(\Phi)$ the mass contained in
the flux tube $\Phi$; $(ii)$ the ratio $b_0$ between the strength of the toroidal and the poloidal components of the magnetic field in the midplane of the source;
$(iii)$ the orientation of the projection of the magnetic axis on the plane of the sky, $\varphi$, measured from north to east (i.e., counterclockwise); and $(iv)$ the inclination with respect to the plane of the sky, $i$, assumed 
to be positive (negative) if the magnetic field in the northern sector points toward (away from) the observer.
The model is isothermal, and the value of 
the sound speed
provides the scaling for both the density and the 
magnetic field strength. In order to match
the observed intensity at $1\arcsec$, we set the effective sound speed to
1.4~km~s$^{-1}$.

We considered three different values for the mass-to-flux ratio, which controls the pinching of the field 
lines as well as the flatness of the density distribution: 
$\lambda=1.63$, corresponding to the case of strong field and flat density profile, $\lambda=8.38$, corresponding to the case of weak field and quasi-spherical density profile, and an 
intermediate case, $\lambda=2.66$ (see Table~1 of Li \& Shu~\cite{li96}).
We also accounted for different values of $b_0$, ranging from the purely poloidal case ($b_0=0$) to the case where the toroidal component is half of the poloidal component in the midplane ($b_0=0.5$). We assumed the temperature profile modeled by 
BEL18, and we verified a posteriori that using a 
constant temperature does not significantly affect our conclusions
(see Sect.~\ref{Bestimate}).

The {\em DustPol} module  is an extension to the Line Modeling Engine
(LIME) radiative transfer code
(Brinch \& Hogerheijde~\cite{brinch10}).
Besides calculating line profiles in the far infrared 
and submillimeter regimes, LIME can ray-trace
given density and temperature profiles, estimating
the continuum flux. {\em DustPol} computes the Stokes
$I$, $Q$, and $U$ maps (see Eqs. 4--8 in Padovani et
al.~\cite{padovani12}) and stores them in FITS format,
which are straightforwardly used as an input for the tasks {\tt simobserve} and
{\tt simanalyze} of CASA, adopting the same antenna configuration of the observing runs.
Finally, from these synthetic maps we generated the polarization angle patterns to be compared with 
those obtained from the observations. 
For each combination of $\lambda$, $b_0$, $\varphi$, and $i$, we performed a chi-squared test for 
the difference between the
observed and the modeled polarization angles 
within the $5\sigma$ contour of the 1.3\,mm dust emission map. We found that the polarization pattern is in general well reproduced by a purely poloidal magnetic field. Adding a small toroidal 
component slightly improves the quality of the fit, provided the latter is not larger than $\sim 10$\% 
of the poloidal component (see Appendix B).
Therefore, from now on, we only consider the case $b_0=0.1$ in our models.
 For the sake of completeness, Appendix~\ref{app:chivsi266} shows the 
results of our modeling for $b_0=0$, 0.25, and 0.5.

Figure~\ref{chi2vsinclphi} shows the reduced chi squared\footnote{The reduced chi squared, $\bar\chi^2$, 
is computed as
$\chi^2/(n-p)$, where $n$ is the number of observed polarization
angles and $p=3$ the number of parameters ($b_0$, $i$, 
and $\varphi$).}, $\bar\chi^2$, as a function of the inclination, $i$, and the orientation, $\varphi$,  for $b_0=0.1$ and the three values of 
$\lambda$. The high values of $\bar\chi^2$ are due to the 
very low average value of the uncertainty on the observed polarization angle, $\delta\psi_{\rm obs}$,
which is less than about $4^\circ$ inside the $5\sigma$ contour
of the 1.3~mm dust continuum emission. 
In contrast, the polarization angle residuals,
$\Delta {\rm \psi=\psi_{\rm obs}-\psi_{\rm mod}}$,
given by the difference between 
the observed (${\rm \psi_{\rm obs}}$) and modeled (${\rm \psi_{\rm mod}}$) polarization angles for the best model, can be as large as $\pm 90^\circ$, because the model does not include the turbulent 
component of the magnetic field (see Sect.~\ref{Bestimate} and Appendix~\ref{app:dpacontours}
for details).

Independently of the value of $\lambda$,
all the curves show a minimum around $\varphi\sim -45^\circ$, but the higher the mass-to-flux ratio, 
the lower is the difference in $\bar\chi^2$ between positive and negative inclinations. 
This is explained by the fact that on the one hand the difference between negative and positive inclinations is significant only for large opacities; on the other hand, a large mass-to-flux ratio corresponds to a less centrally-condensed source,
which is more optically thin. We verified this statement using {\em DustPol}, which also calculates the dust opacity, $\tau$, and the values obtained for mass-to-flux ratios of 1.63, 2.66, and 8.38 at the density peak are 1.1, 0.8, and 0.6, respectively.
This allowed us to discard the case $\lambda=8.38$, since we know that G31 is
optically thick toward the center (BEL18). 
We found that the best model is given by $\lambda=2.66$, $i=-{45^\circ}_{-4^\circ}^{+3^\circ}$, and $\varphi=-{44^\circ}_{-4^\circ}^{+6^\circ}$. The $1\sigma$ errors have been estimated
using the method of Lampton et al.~(\cite{lampton76}).

\begin{figure}
\centerline{\includegraphics[angle=0,width=8.5cm,angle=0]{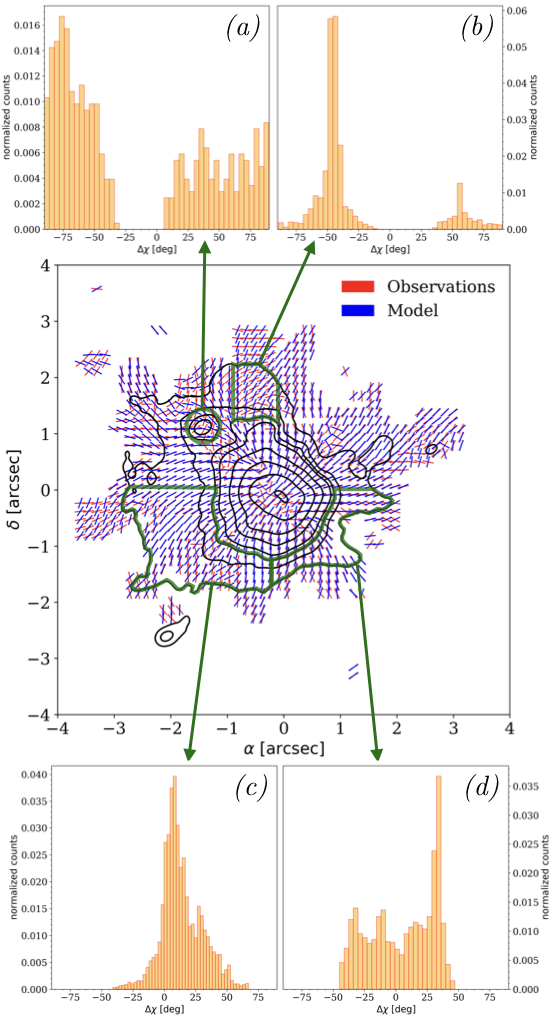}}
\caption{{\it Middle plot}: comparison between the observed and the modeled polarization angles 
superposed to the contours of the 1.3~mm dust continuum emission (see Fig.~\ref{vectormap} for details).
 The four histograms show the polarization angle residuals for the NE core ({\it a}), and for the regions associated with two of the outflows mapped in SiO in the core: that associated with the  N--S outflow located to the north of the Main core ({\it b}), and those associated with the E--W outflow to the southeast ({\it c}) and southwest ({\it d}) of the Main core (see Sect.~\ref{sect:perturb}). For the latter regions ({\it b, c}, and {\it d}), the histograms of the polarization angle residuals have been calculated in the regions encompassing the $5\sigma$ and $15\sigma$ contours.}
\label{histoappendix}
\end{figure}

Figure~\ref{vectormap} shows the comparison between the polarization angles obtained from
observations and those from the best fit model. Even if the overall result is consistent with our best model,  we remark that there 
are a few regions showing departures from this geometry (see Fig.~\ref{histoappendix}): 
$(i)$ the center of the Main core
where there are four embedded sources (see Fig.~\ref{fig-cont}) and we may expect a 
more complex configuration of the magnetic field lines; 
$(ii)$ the NE core, where the magnetic field appears to show an
independent poloidal configuration inside a radius of $\lesssim0.5\arcsec$
probably due to its own collapse; and $(iii)$ the northern and southwestern edges of the Main core, where  two different molecular outflows have been traced in SiO (BEL18). See Sect.~\ref{sect:disc} for a detailed discussion on the possible causes of the deviation.  

\subsection{Magnetic field strength estimate}
\label{Bestimate}

Now, we take advantage of our model fit to measure the magnetic field strength using the Davis-Chandrasekhar-Fermi method (Davis~\cite{davis51}; Chandrasekhar \& Fermi~\cite{chandra53}). For this purpose, it is necessary to estimate the dispersion of the polarization angle with respect to the model. We thus computed the histogram of the polarization angle residuals. We considered the region inside the $15\sigma$ contour level to focus only on the Main core, avoiding any 
contamination by the NE core.
The average uncertainty on the observed polarization angles is 
$\delta\psi_{\rm obs}=0.5~\sigma_{QU}/\sqrt{Q^2+U^2}$, 
where $\sigma_{QU}=22~\mu$Jy\,beam$^{-1}$ is the noise on the observed Stokes\,$Q$ and $U$. This expression is valid for high ($>5$) signal-to-noise ratios (e.g., Vaillancourt~\cite{vaillancourt06}), which is the case of our data. As seen in Fig.~\ref{fig-pol} ({\it top panel}), the linearly polarized intensity, $P$, of the G31 core is $>5\sigma_{QU}$. In the region inside the $15\sigma$ contour
of the 1.3~mm dust continuum map, 
$\delta{\rm \psi_{\rm obs}}\lesssim 2^\circ$ (see Fig.~\ref{dpacontours}), therefore, we used a
histogram bin of $4^\circ$.

\begin{figure}[!h]
\centerline{\includegraphics[angle=0,width=8.5cm,angle=0]{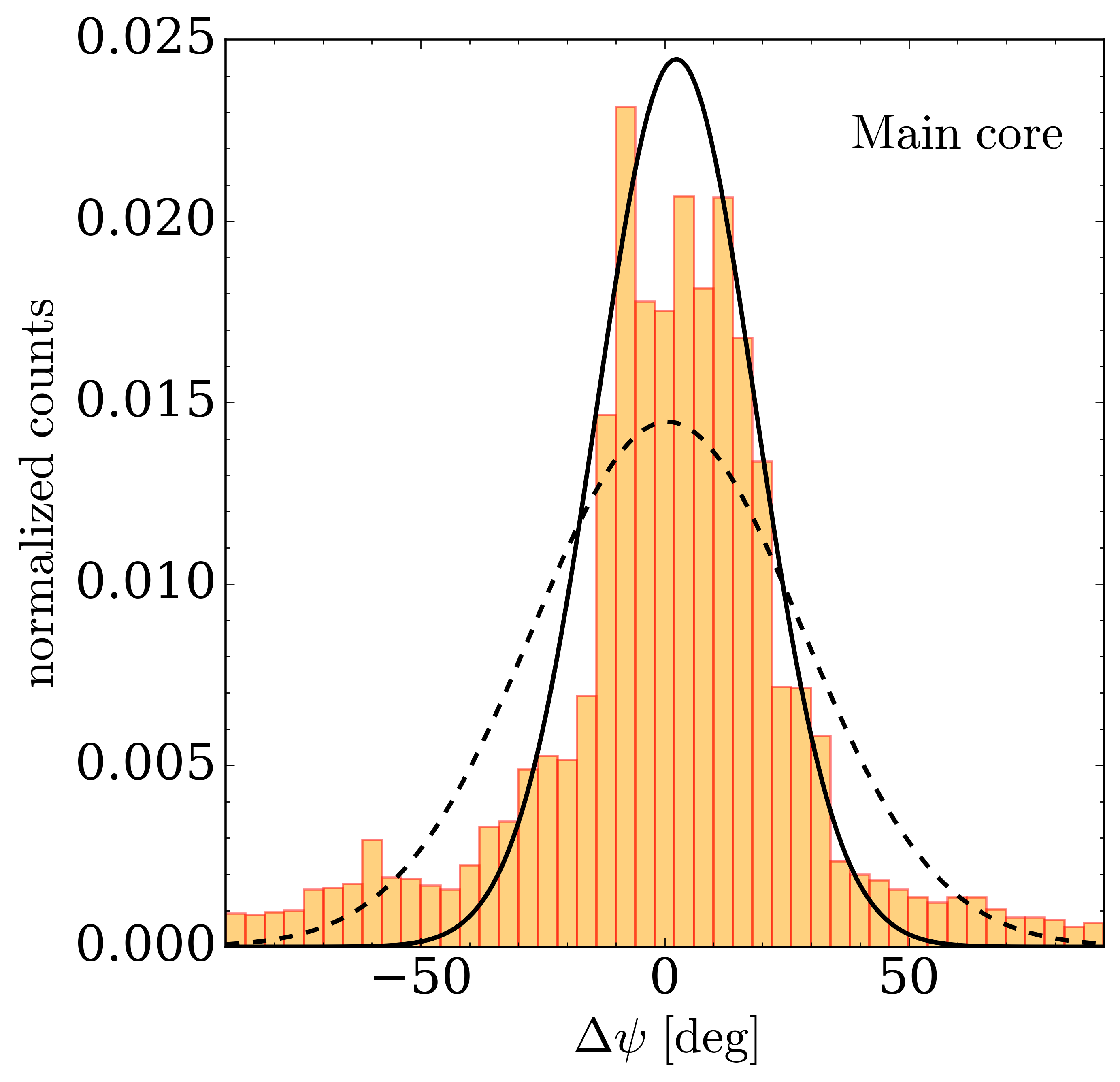}}
\caption{Histogram of the polarization angle residuals for our best model in the region inside the $15\sigma$ contour (corresponding to the Main core) of the 1.3~mm dust continuum map. The dashed and solid black lines show two Gaussian fits obtained considering the whole range of $\Delta\psi$ and limited to
$\pm 45^\circ$, respectively.}
\label{histog31}
\end{figure}

\begin{figure}
\centerline{\includegraphics[angle=0,width=9.5cm,angle=0]{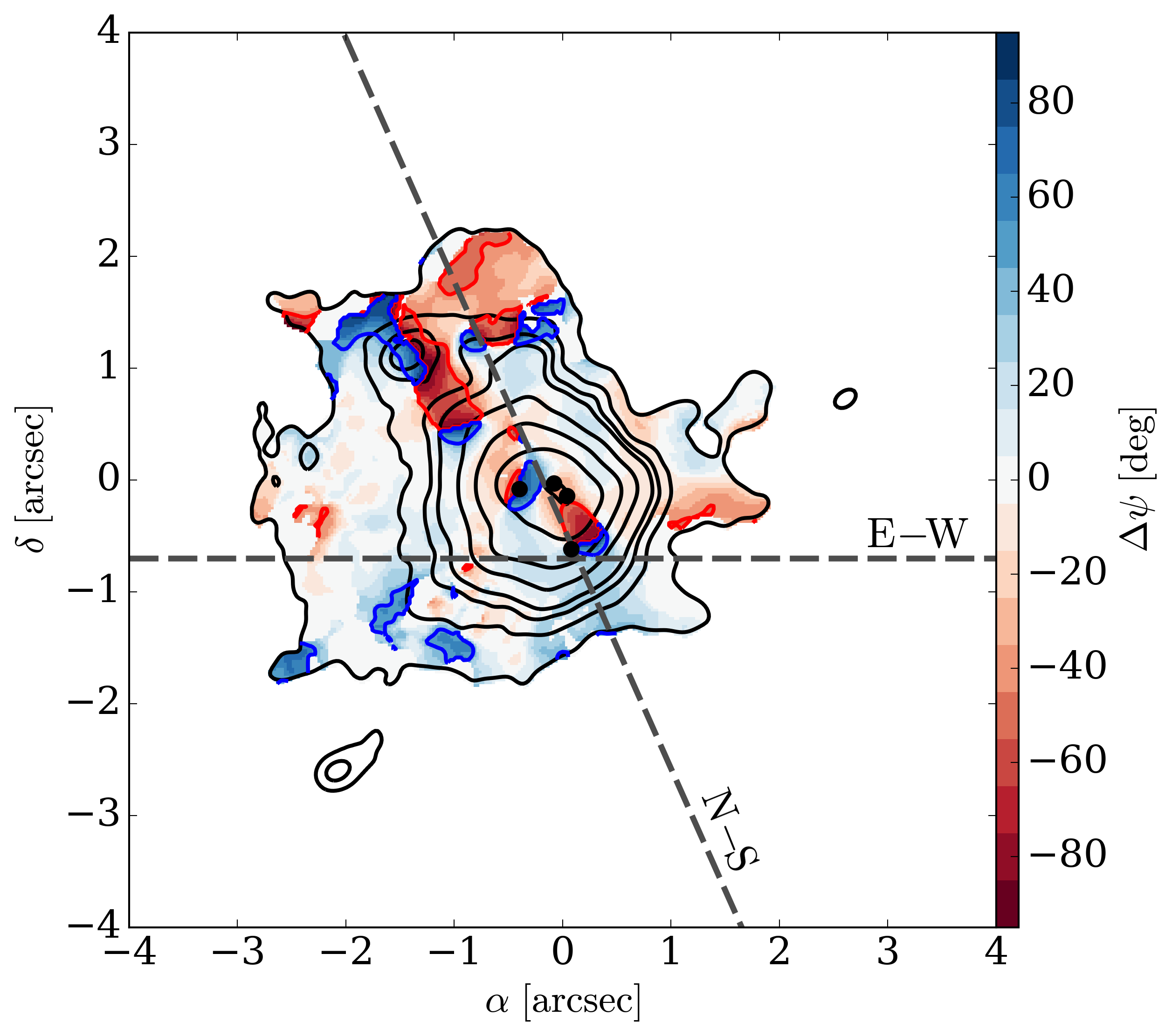}}
\caption{Polarization angle residuals ({\it color map}) superposed the 1.3 mm dust continuum emission contours ({\it gray solid lines}). The black dots mark the position of four embedded continuum sources (see Fig.~\ref{fig-cont}).
Blue and red contours show the regions where
$\Delta\psi>45^\circ$ and $<-45^\circ$, respectively.
Dashed gray lines indicate the direction of the E--W and N--S outflows mapped in SiO by BEL18.}
\label{errorplot_b001}
\end{figure}

Figure~\ref{histog31} shows the distribution of polarization angle residuals for our best model. A Gaussian fit over the whole
distribution of residuals gives an average value of 
$0.6^\circ \pm 27.6^\circ$.  As shown in this figure, the polarization angle residuals can be very large, up to  $\pm 90^\circ$. To better discriminate the areas with large dispersion, we plotted in Fig.~\ref{errorplot_b001} the polarization angle residuals over the whole G31 core. This figure shows that the residuals $\Delta \psi$ in the Main core (inside the 15$\sigma$ contour level) are between $-45^\circ$ and $45^\circ$ except for small areas associated with the four embedded sources or in the direction of the NE core and the north--south (N--S) outflow mapped in SiO by BEL18.  As already mentioned in the previous section, these regions will be analyzed in detail in Sect.~\ref{sect:disc} and the possible causes of the large deviations from the model will be discussed. Therefore, assuming that these high dispersion values indicate areas where the initial magnetic field associated with the Main core has been disturbed, we decided to limit the range of $\Delta\psi$ between $-45^\circ$ and $+45^\circ$. In this case, the average value of the distribution of residuals is $2.3^\circ \pm 17.3^\circ$.  Therefore, the standard deviations on the polarization angle dispersion from the two Gaussian fits,
$\sigma_\psi$, are $27.6^\circ$ and $17.3^\circ$. If in our model 
we adopt a constant temperature of 250~K, which is the mean value in the G31 Main core (BEL18), the average value of the distribution of residuals is $2.6^\circ\pm30.2^\circ$, considering the whole angle distribution, 
and $5.1^\circ\pm16.9^\circ$, if limited in the range $\pm45^\circ$. Therefore, 
the difference between the average values of 
the distributions of residuals in the isothermal and non-isothermal cases is not significant. Since the measurement uncertainty of the polarization angle $\delta\psi_{\rm obs}$ is $\lesssim 2^\circ$, 
the intrinsic dispersion is $\delta\psi_{\rm int}$ = $(\sigma_\psi^2 -  \delta\psi_{\rm obs}^2)^{1/2}\sim \sigma_\psi$.

To estimate the strength of the magnetic field on the plane of the sky, $B_{\rm pos}$, we used the
Davis-Chandrasekhar-Fermi method, which is based on the 
assumption that the perturbations responsible for the polarization
angle dispersion $\delta\psi_{\rm int}$ are Alfv\'en waves of amplitude $\delta B_{\rm los}=\sqrt{4\pi\rho}\,\sigma_{\rm los}$, where $\rho$ is the density and $\sigma_{\rm los}$ is the line-of-sight velocity dispersion. This gives
\be
B_{\rm pos}=\xi\frac{\sigma_{\rm los}}{\delta\psi_{\rm int}}\sqrt{4\pi\rho}\,,
\ee
where $\xi=0.5$ is a correction factor derived from turbulent cloud simulations (Ostriker et al.~\cite{ostriker01}). The line-of-sight velocity dispersion $\sigma_{\rm los}$ has been computed from the line width $\Delta V$ of the CH$_3$CN observations of BEL18, which have an angular resolution similar to that of our polarization observations, as $\sigma_{\rm los}$ = $\Delta V/\sqrt{8\,\ln{2}}$. To avoid the effects of rotation on the line broadening,  we estimated $\Delta V$ at different pixel positions of the core and then averaged the values. The thermal contribution to the velocity dispersion is negligible for the temperatures $\ga 100$\,K estimated for the Main core (e.g., BEL18). Using $\Delta V\simeq5$\,\kms, the value of $\sigma_{\rm los}$ is 2.1\,\kms. This value is consistent with the turbulent velocity dispersion of  2.7\,\kms\ obtained by Osorio et al.~(\cite{osorio09}) from the modeling of the G31 core. The average volume density of the Main core has been computed assuming spherical symmetry from the mass estimate of BEL18, which has been computed for a radius, $R$, of $1\farcs1$. This radius corresponds to that of the Main core inside the $15\sigma$ contour level (see Fig.~\ref{fig-cont}). We note that the mass of 120\,$M_\sun$ of BEL18 was estimated for a distance of 7.9\,kpc. Assuming a distance of 3.7\,kpc, the mass of the Main core is $\sim$26\,$M_\sun$ and the corresponding mean number density is $n=1.4\times 10^7$~cm$^{-3}$. Our modeling provides the inclination with respect to the plane of the sky, $i$, which allows us to estimate the total magnetic field strength as $B=B_{\rm pos}/\cos i \simeq8$--13\,mG, for an intrinsic dispersion of $27.6^\circ$ and $17.3^\circ$, respectively. GIR09 have estimated a value of $B_{\rm pos}$ of $\sim 14$\,mG (corrected for a distance of 3.7\,kpc), that for the inclination angle of $-45^\circ$ derived from our modeling, corresponds to a strength of the magnetic field of $\sim 20$\,mG, slightly higher than our estimate. This result is consistent with the range of magnetic field strengths predicted by the model 
(12\,mG at $0\farcs2$ and 3\,mG at $1\arcsec$, see
Fig.~\ref{fieldlines}).

Finally, we evaluated the mass-to-flux ratio from Eq.~(\ref{m2fratio}) to check the consistency of our model (with $\lambda > 1$, ``supercritical'') with the observations. We computed the magnetic flux, $\Phi=\pi R^2 B$, inside
a radius of $1\farcs1$, corresponding to the Main core region, assuming spherical symmetry, and obtained a range of $(0.9$--$1.5)\times10^{32}$\,G\,cm$^2$. As for $M(\Phi)$, the mass contained in the flux tube $\Phi$, we assumed that this is the mass of the core (26\,$M_\sun$) plus the mass of the (proto)stars already formed in the core. We estimated the mass of the (proto)stars from the bolometric luminosity of the region, which is $\sim4.4\times10^4$\,$L_\sun$. If we assume that the bolometric luminosity of the core is mainly produced by a single main-sequence star, then the mass of such star would be of $\sim$20\,$M_\sun$, corresponding to an O8.5--O9 star (Mottram et al.~\cite{mottram11}). Consequently, $M(\Phi)=46$\,$M_\sun$ and $\lambda$ lies in the range 1.0--1.6. The definition of $\lambda$ in Eq.~(\ref{m2fratio}) accounts for the mass 
inside a flux tube, while from observations we estimated the mass in a sphere of radius $1.1\arcsec$. We then multiply the latter by a correcting factor equal
to 1.4 (Li \& Shu~\cite{li96}) to obtain $\lambda\sim 1.4$--$2.2$, slightly supercritical. We note that the observed value of the mass-to-flux ratio should be taken as a lower limit because the mass of the core should be taken as a lower limit and the stellar mass content could be larger if the luminosity of 4.4$\times10^4$\,$L_\odot$
 originates from multiple stars instead of a single star as previously assumed by us. In addition, Estalella et al.~(\cite{estalella19}) have recently modeled the line-of-sight velocity of ammonia as a function of projected distance in the G31 core and have obtained a central mass  $\ga$44\,$M_\odot$. Therefore, we believe that the mass-to-flux ratio computed from the observations is consistent with $\lambda=2.66$ assumed for the model.

\section{Discussion}
\label{sect:disc}

\subsection{A supercritical core}

The mass-to-flux ratio of $\lambda>1.4$ obtained by us with a beam of $0\farcs2$ is consistent with the value of $\sim$1.8 estimated by GIR09 with $1''$ resolution, after correcting for the new 3.7\,kpc distance. We note, however, that Girart et al.\ used $B_{\rm pos}$ to estimate $\lambda$. Using the value of $B$ (assuming an inclination angle of $-45^\circ$), the mass-to-flux ratio estimated from the 870\,$\mu$m dust emission would be 1.3. The estimated mass-to-flux ratio suggests that the G31 core is slightly supercritical at different scales, from $\sim$800 to $\sim$4000\,au. This is further supported by the detection of signatures of infall, such as red-shifted absorption and a central spot of blue-shifted emission (GIR09; Mayen-Gijon et al.~\cite{mayen14}; BEL18; Estalella et al.~\cite{estalella19}), and by the presence of at least four embedded sources in the Main core. We estimated the Alfv\'en velocity from the expression\footnote{We note that there is a typo in the formula of GIR09. The Alfv\'en velocity should be proportional to $B$, not to $\sqrt{B}$. The value of $v_{\rm A}$ has been correctly estimated by GIR09.} $v_{\rm A}=B/\sqrt{4\pi\rho}$. Using the estimates of the magnetic field strength in G31, we obtain an Alfv\'en velocity in the range 3--5\,\kms. These values are comparable to, or slightly lower than, the infall velocities estimated by BEL18 from red-shifted absorption ($\sim 2$--$8$\,\kms). The highest infall velocities have been estimated for the vibrationally excited transitions of CH$_3$CN and for some transitions of the isotopologues $^{13}$CH$_3$CN and CH$_3^{13}$CN, which are optically thinner and trace material close to the central (proto)star(s). Therefore, this suggests that while the collapse in the external part of the core is slightly sub-Alfv\'enic, it becomes super-Alfv\'enic close to the center.

The Davis-Chandrasekhar-Fermi method also allows to estimate the ratio of the turbulent component of the 
magnetic field, $\delta B\sim \sqrt{3}\,\delta B_{\rm los}$, to the uniform 
component $B=B_{\rm pos}/\cos i$. With $B_{\rm pos}$ given by Eq.~(2), this ratio results
\begin{equation}
\frac{\delta B}{B}\sim \frac{\sqrt{3}\cos i}{\xi}\delta\psi_{\rm int}.
\end{equation}
Inserting the numerical values, we obtain $\delta B/B\sim0.7$ and
$\sim1$, for  $\delta\psi= 17.3^\circ$ and $27.6^\circ$, respectively. This indicates
that the energy of the turbulent component of the
magnetic field in the Main core of G31 is a significant fraction
(50\% or larger) of the energy of the uniform component of the field
included in our model.

\begin{figure}
\centerline{\includegraphics[angle=0,width=8.5cm,angle=0]{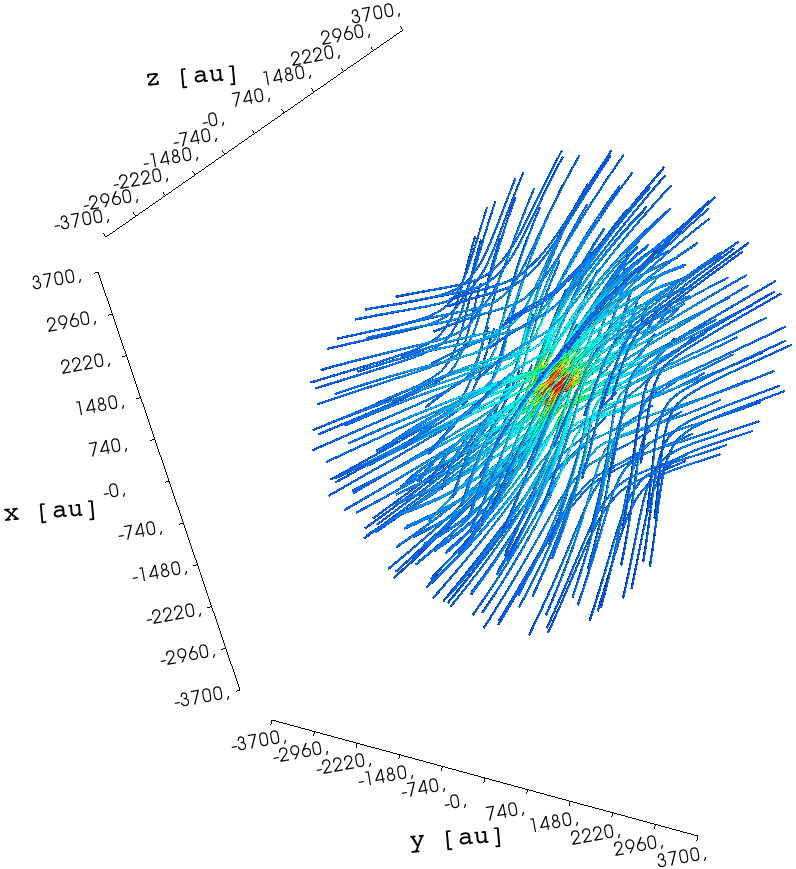}}
\vspace*{.8cm}
\centerline{\includegraphics[angle=0,width=8.5cm,angle=0]{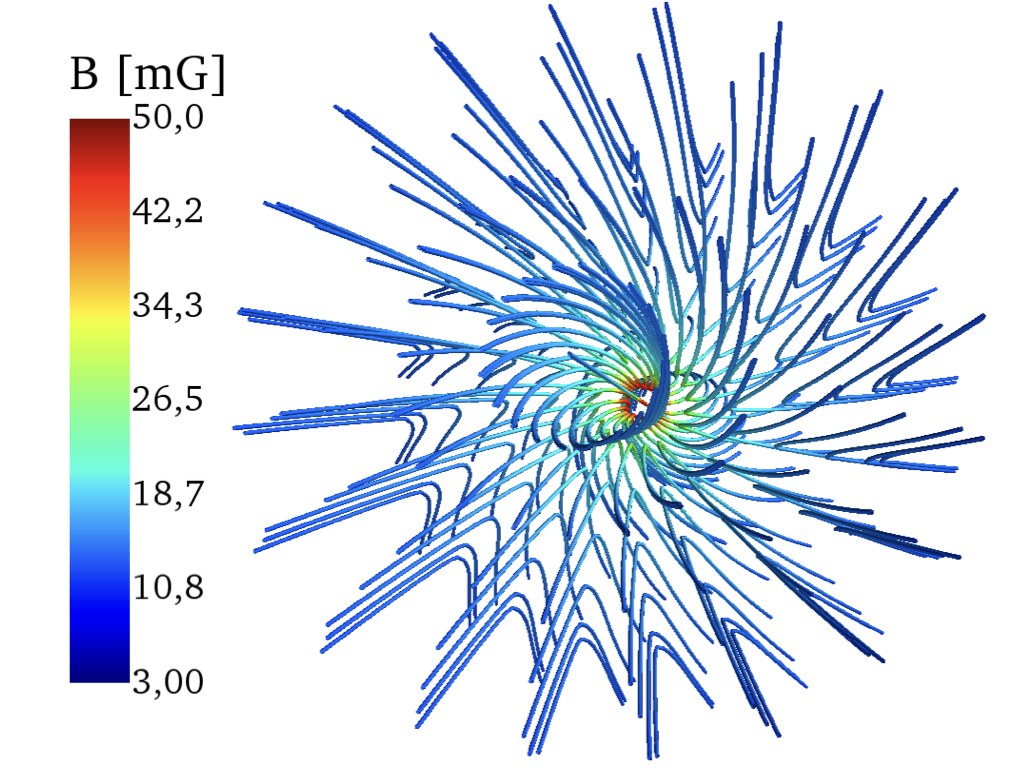}}
\caption{{\it Upper panel}: magnetic field configuration inside a radius of 3700~au, corresponding to the radius of the Main core in G31, for $b_0=0.1$, $i=-45^\circ$, and $\varphi=-44^\circ$. {\it Lower panel}: magnetic field line configuration for $b_0=0.1$ shown almost pole-on to emphasize the line twisting close to the symmetry axis. The spatial scale is the same in both panels (the model extends up to a radius of 3700\,au). The color of the magnetic field lines indicates the magnetic field strength in both panels (see color scale in the bottom panel).}
\label{fieldlines}
\end{figure}

\subsection{Confirming the rotating toroid model}

Molecular line observations have revealed a striking northeast-southwest (NE--SW) velocity gradient in the G31 core on scales from 10$^3$--10$^4$\,au (Beltr\'an et al.~\cite{beltran04}, \cite{beltran05}, BEL18; Araya et al.~\cite{araya08}; GIR09; Cesaroni et al.~\cite{cesa11}).
The interpretation of this NE--SW velocity gradient has been long controversial. Some authors (Beltr\'an et al.~\cite{beltran04}, \cite{beltran05}, BEL18; GIR09; Cesaroni et al.~\cite{cesa11}) propose that the velocity gradient is produced by the rotation of the core. In contrast, Araya et al.~(\cite{araya08}) interpret such a gradient as due to a compact bipolar outflow. Our model fit allows us for the first time to discriminate between both scenarios. 

 The position angle of the magnetic axis, $\varphi$, is probably the quantity better constrained by the fitting procedure, because as already mentioned in Sect.~\ref{sect:model}, independent of the mass-to-flux ratio of the model, the symmetry axis of the hourglass B field is oriented SE--NW, with $\varphi\sim -45^\circ$.  This orientation is almost perpendicular to the NE--SW velocity gradient and to the main axis of the core dust continuum emission (BEL18). In addition, as shown in Fig.~\ref{chi2vsinclphi}, the chi-squared function is maximum at $\sim 40^\circ$, which suggests that this direction is the most unlikely for the magnetic field. This result points to the fact that the magnetic field is oriented perpendicular to the plane of the core and almost parallel to the rotation axis. This is important because it supports the hypothesis that the velocity gradient is due to rotation and discards the molecular outflow scenario proposed by Araya et al.~(\cite{araya08}).

\subsection{Influence of rotation on the magnetic field}

The angular velocity $\Omega$ of the rotation of the G31 core has been estimated by BEL18 for different transitions with different energies of CH$_3$CN and its isotopologues and is $\Omega=(2.7$--$5.4)\times10^{-12}$\,s$^{-1}$ for a radius of 3700\,au, depending on the transition. This rotation appears to have little effect on the magnetic field, as suggested by the modeling (see Fig.~\ref{fieldlines}). In fact, as seen in Sect.~\ref{sect:model}, we find that the magnetic field geometry in G31 is well represented by a toroidal component not larger than 10\% of the poloidal component. A similar behavior has been observed in the low-mass systems BHB07-11 (Alves et al.~\cite{alves18}) and VLA\,1623-A in Ophiuchus (Savadoy et al.~\cite{savadoy18}), despite the different spatial scales and associated protostellar masses. In any case, we stress that, as shown in Fig.~\ref{fieldlines}, the direction of rotation of the modeled toroidal component coincides with that of the core, which rotates clockwise.

A possible explanation for the fact that rotation does not seem to affect the magnetic field could be a decoupling of the envelope and the magnetic field. A partial decoupling could be produced by several 
causes, for example: the removal by coagulation of very small charged grains, with size $\sim 10^{-2}$~$\mu$m, that dominate the coupling of the bulk neutral matter to the magnetic field (Zhao et al.~\cite{zhao16}); or an attenuation of 
the flux of ionizing cosmic-ray particles expected in the dense regions around a forming protostar (Padovani et al.~\cite{padovani14, padovani18}). Another possibility to explain such a small toroidal component could be that the core is still young and therefore rotation had no time to affect the magnetic field yet.  Starting from a purely poloidal magnetic field, a rotation of the core of an angle of $\sim \pi/2$, occurring on a timescale on the order of  $\pi/(2\Omega)=(1.0$--$2.0)\times 10^4$~yr, would generate in the source's midplane a field with toroidal-to-poloidal intensity ratio $b_0\sim 1$. Thus, if the magnetic field and the envelope are well coupled, the value of $b_0\lesssim 0.1$ resulting from our modeling would imply an ``age'' of a few $10^3$~yr, suggesting that the G31 core is very young.

\subsection{Perturbation of the magnetic field}
\label{sect:perturb}

The polarized emission in G31 has been successfully fit with a semi-analytical magnetostatic model of a toroid supported by magnetic fields. As seen in the previous section, the magnetic field associated with G31 Main core is basically poloidal. This almost poloidal model fits  the observed magnetic field lines in most of the core extremely well but it deviates from the observations in some areas. Figures~\ref{histoappendix} and \ref{errorplot_b001}
show that our model fails to properly describe the magnetic field close to the center of the Main core, around the NE core, and at the northern and southwestern edges of the Main core. In the next sections, we discuss the possible causes of the large deviations from the model in these areas.

\subsubsection{The central region} 

As shown in Fig.~\ref{errorplot_b001}, the discrepancies between the observed and predicted polarization angles are large close to the center of the Main core, and in particular at the position of the four embedded sources (see Sect.~\ref{sect:cont}). The dispersion in the residuals in this area is large ($|\Delta \psi|>45^\circ$), compared to the average values in the Main core. The presence of the embedded sources indicates that fragmentation has already taken place in the core. Therefore, this suggests that, despite its strength ($\sim$10\,mG), the magnetic field is not sufficient to prevent the fragmentation and collapse of the core.

The embedded sources appear to be in the accretion phase, as suggested by the detection of red-shifted absorption toward them (Beltr\'an et al., in prep.). Therefore, the existence of different collapsing centers that might drag and perturb the larger-scale magnetic field makes it very difficult to model the magnetic field in such environment with our idealized model. This could explain the large polarization angles residuals toward the embedded sources at the center of the Main core.

\begin{figure*}
\centerline{\includegraphics[angle=0,width=18cm,angle=0]{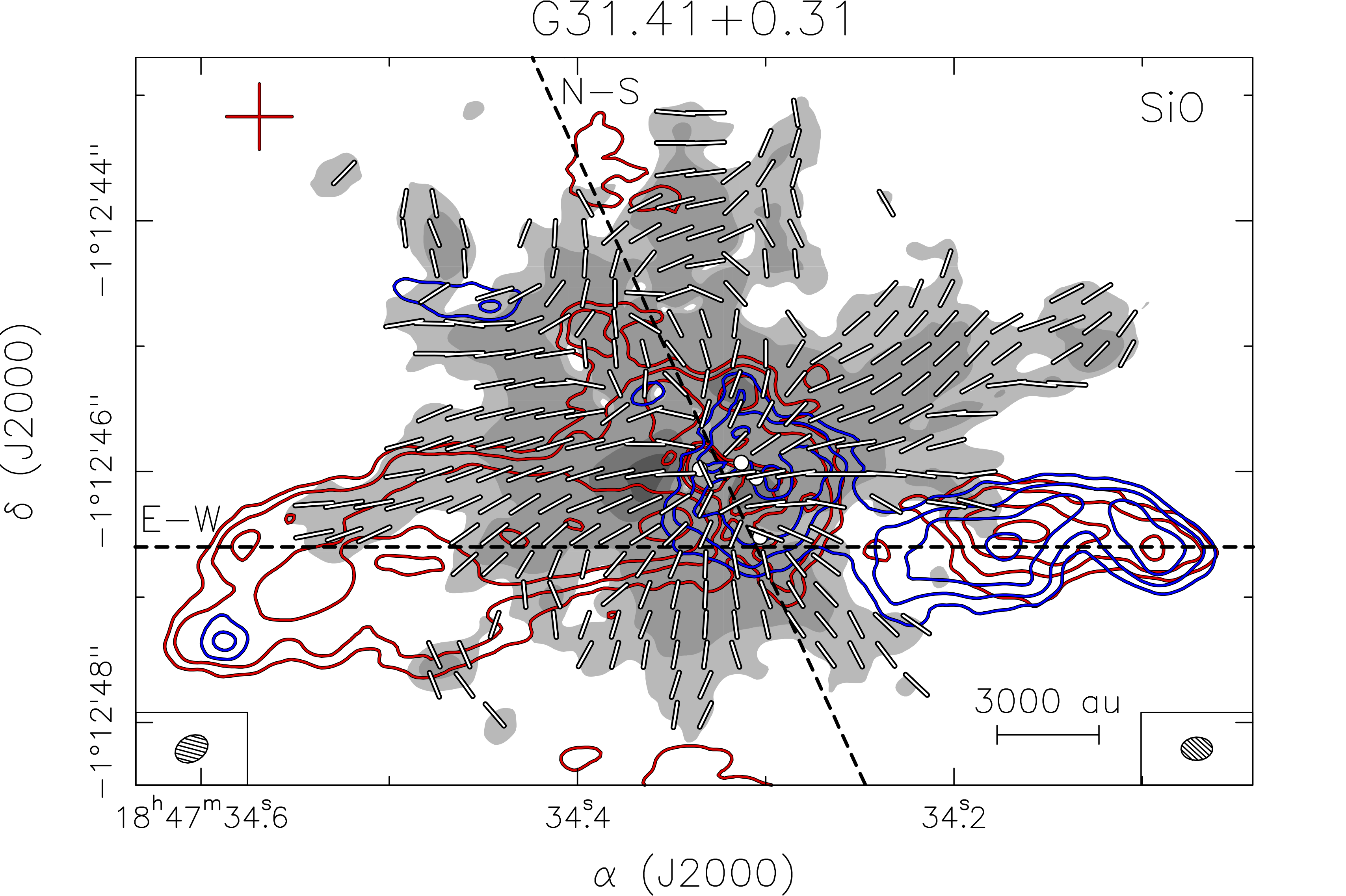}}
\caption{Overlay of the blue-shifted ({\it blue contours}) and red-shifted ({\it red contours}) SiO\,(5--4) averaged emission from BEL18 on the polarized intensity map ({\it gray scale}). The blue-shifted and red-shifted emission have been averaged over the (73, 90)\,\kms\ 
and (103, 119)\,\kms\ velocity interval, respectively. Contour levels are 3, 6,
12, and 24 times 1.1\,mJy\,beam$^{-1}$. Gray-scale contours for the polarized intensity are 5, 10, 50, and 90 times $\sigma$, where $1\sigma$ is 22\,$\mu$Jy\,beam$^{-1}$. The white segments, plotted every ten pixels, indicate the magnetic field lines. White dots mark the position of the four embedded continuum sources (Beltr\'an et al., in prep.). Dashed black lines indicate the direction of the E--W and N--S outflows mapped in SiO by BEL18. The synthesized beam of the polarized emission and of the SiO observations are shown in the lower left-hand and lower right-hand corner, respectively. }
\label{fig-outflow}
\end{figure*}

\subsubsection{The NE core} 

Albeit smaller and weaker than the Main core, the NE core is also a HMC, as indicated by the detection of several transitions of methyl cyanide and methyl formate (see Fig.~A.1 of BEL18).  The magnetic field toward this core is strongly disturbed and cannot be properly fit with our model (see Fig.~\ref{histoappendix}). The polarization angle residuals are large ($|\Delta \psi| > 45^\circ$) (Fig.~\ref{errorplot_b001}). As shown in Fig.~\ref{fig-cont}, the observed magnetic field at the position of the NE core, in particular to the northeast of the core, appears to show an independent poloidal configuration (hourglass shape) that could be produced by gravitational collapse. If we assume a dust temperature of 50--100\,K, consistent with the fact that the NE core is a HMC, we obtain a mass of the core $M_{\rm NE}\simeq3$--6.5\,$M_\odot$, for an  integrated flux density of 52\,mJy (BEL18), a dust absorption coefficient per unit mass $\kappa_\nu=0.008$\,cm$^2$g$^{-1}$ at 217\,GHz 
(Ossenkopf \& Henning~\cite{ossenkopf94}), and a gas-to-dust mass ratio of 100.

In this scenario, the magnetic field lines are dragged to the center of collapse associated with the NE core and perturb the magnetic field associated with the Main core. In our model, the magnetic field is associated and centered with the Main core, and therefore, we cannot reproduce the perturbations resulting from the collapse of the NE core. 

The large-scale magnetic field, to the north and to the east of the position of the NE core (see Figs. \ref{vectormap} and \ref{histoappendix}), also follows an hourglass morphology which could indicate the presence of a filament channeling mass inflow to the center of the G31 core where NE and Main are embedded. 

\subsubsection{The molecular outflows}

The 870\,$\mu$m SMA polarized observations by GIR09 revealed a clear lack of polarized emission to the west of the core dust continuum emission peak (see their Fig.~1A). Our ALMA observations have confirmed this lack of polarized emission, and have better pinpointed its location at the southwestern edge of the Main core (Fig.~\ref{fig-pol}; {\it top panel}). This unpolarized region appears to coincide with the blue-shifted, western lobe of an SiO molecular outflow (the E--W outflow mapped by BEL18; see Fig.~\ref{fig-outflow}). This outflow is centered $\sim$0$\farcs6$ to the south of the dust continuum emission peak and could be associated with the southernmost of the four embedded sources.  The polarized emission also seems to surround the red-shifted lobe of this outflow to the east (Fig.~\ref{fig-outflow}).
In both cases, the polarized emission traces the walls of the cavities opened by the outflow. The fact that there is no polarized emission along most of the molecular outflow could be due to the lack of dust, which might have been evacuated by the outflow. This has also been observed in low- and intermediate-mass protostars (e.g., Serpens SMM1: Hull et al.~\cite{hull17b}; B335: Maury et al.~\cite{maury18}). In some cases the polarization is enhanced along the walls of the outflow cavity (e.g.,  B335: Maury et al.~\cite{maury18}; Ser-emb 8(N): Hull et al., in prep.). This polarization enhancement is also marginally observed in G31 (see Fig.~\ref{fig-pol}).  All this has led to suggestions that the outflow might shape the magnetic field (Hull et al.~\cite{hull17b}). One expects that the outflow sweeps away the core material, thereby creating heated and compressed regions at its edges, and it is in these regions that the magnetic field is disturbed. 

In fact, this effect is observed in G31. The magnetic field at the base and along the walls of the E--W outflow cavities is partially perturbed. As shown in Fig.~\ref{errorplot_b001}, $|\Delta \psi| < 30^\circ$ except for localized regions. This is especially true for the northern part of the blue-shifted, western lobe and the southern part of the red-shifted, eastern lobe.  We speculate that the perturbed field is detected in regions where there is no significant dust emission from the core along the line-of-sight or this component has been filtered out by the interferometer.

The other effect of the molecular outflow is that, by generating an almost dust free cavity, the photons from the inner regions around the protostar can more easily escape from the core, illuminating the cavity. This could enhance the polarization efficiency through radiative torques (Hoang \& Lazarian~\cite{hoang09}, Andersson et al.~\cite{andersson15}). There is some evidence of higher polarization degree in the northern part of the blue-shifted, western lobe. In addition, the red-shifted eastern lobe shows polarization where no Stokes\,$I$ emission is detected, which also may indicate an enhancement of the polarization. That only Stokes\,$Q$ and $U$ emission is detected in the red-shifted eastern lobe is due to the sensitivity of the maps, because while the rms noise of the maps for Stokes\,$Q$ and $U$ is 22\,$\mu$Jy\,beam$^{-1}$, that for Stokes\,$I$ is a factor 55 higher (1.2\,mJy\,beam$^{-1}$). 

BEL18 have also mapped a N--S SiO molecular outflow in the region, which is hardly visible in Fig.~\ref{fig-outflow}. This outflow, which is much weaker than the  east-west one, could be driven by one of the embedded centimeter continuum sources detected by  Cesaroni et al.~(\cite{cesa10}) at the center of the Main core. Despite the fact that this outflow has not yet excavated wide cavities, it appears to have strongly disturbed the magnetic field at the northern edge of the Main core. As shown in Figs.~\ref{histoappendix} and \ref{errorplot_b001}, the observed and modeled polarization angles strongly diverge to the north and the polarization angle residuals are always larger than $30^\circ$.

\subsection{Comparison with other regions}

The direct {\it face to face} comparison between the dust polarization pattern and theoretical predictions of the collapse of a magnetized dense core can basically be done when the observed magnetic field morphology resembles relatively simple magnetic field poloidal, toroidal configuration, or a combination of them (e.g.,  Gon\c{c}alves et al.~\cite{goncalves08}; Alves et al.~\cite{alves18}). This is the case for the two well known low- and high-mass star-forming cores, G31 and NGC\,1333 IRAS\,4A (Frau et al.~\cite{frau11} and this work). Interestingly, the two cores present a velocity gradient almost perpendicular to the main direction of the magnetic field (Beltr\'an et al.~\cite{beltran05}; Ching et al. \cite{ching16}), but the field lines appear not to be perturbed by the rotation. Of these two cores, G31 appears to have the  mass-to-flux ratio closest to the critical value. 

The hourglass morphology has also been reported in other massive cores  (Schleuning et al.~\cite{schleuning98}; Qiu et al.~\cite{qiu14}; Li et al.~\cite{li2015}), and in low-mass cores (Girart et al. \cite{girart99}, \cite{girart06}; Davidson et al.~\cite{davidson14}; Kown et al.~\cite{kwon18}; Maury et al.~\cite{maury18}). These are clear cases of magnetically regulated star formation with a relatively uniform magnetic field at core and/or envelope scales.  However, there are other cases reported in the literature that 
indicate the opposite, namely that the magnetic field plays a minor role in the star-formation process. In some cases, the magnetic field appears to be affected by the stellar feedback  (Tang et al.~\cite{tang09}, \cite{tang10}; Frau et al.~\cite{frau14}), but in other cases, the magnetic field appears to be weak energetically with respect to turbulence, angular momentum and/or gravity (Girart et al.~\cite{girart13};  Hull et al.~\cite{hull17a}; Cortes et al.~\cite{cortes16}; Ju\'arez et al.~\cite{juarez17}). This suggests that there is a diverse initial condition at the onset of gravitational collapse. Statistically, polarization observations with single-dish telescopes of a significant sample of star-forming clouds found a  significant fraction of sources for which the magnetic field appears to be relevant  (Koch et al.~\cite{koch14}). Probing similar spatial scales as in this study of G31, Zhang et al.~(\cite{zhang14}) presented polarization studies of 14 massive molecular clumps with the SMA. By comparing with magnetic field orientations at the parsec scale, they found  that the field in dense cores is correlated with that in their parental clumps, thus, concluded on a statistical basis that magnetic fields  are dynamically important in shaping the fragmentation of the parsec-scale clumps and the formation of dense molecular cores. Furthermore, they found no strong correlation between the core magnetic field orientation and the major axis of molecular outflows (see also Hull et al. \cite{hull13}; Galametz et al.~\cite{galametz18}), which suggests that the role of magnetic fields are weakened relative to gravity and angular momentum at scales from cores to accretion disks. In comparison to these statistical studies, G31 represents a case that magnetic fields are dynamically dominant at scales of dense cores, and maintain the importance down to the scale of $10^3$ au. As discussed in Hull \& Zhang~(\cite{hull19}), this hourglass-shaped magnetic field configuration is relatively rare in both high-mass and low-mass star forming cores.

\section{Conclusions}

We carried out ALMA 1.3\,mm high-angular ($\sim$0$\farcs2$) resolution polarization observations of the hot molecular core G31.41+0.31, previously observed with the SMA at 870\,$\mu$m with lower ($\sim$1$''$) angular resolution. The ALMA observations have confirmed the hourglass-shaped magnetic field morphology observed previously with the SMA.  

The polarization fraction in the HMC ranges  from 0.1\% to 13\%. However,  these values should be taken as lower limits due to the fact that at least $\sim$40\% of the total flux measured for Stokes\,$I$ is contaminated by line emission.

The polarized emission in the central region of G31 has been successfully fit with a semi-analytical magnetostatic model of a toroid supported by magnetic fields. The best fit model suggests that the magnetic field associated with the Main core in G31 is well represented by a purely poloidal field, with a possible hint of a toroidal component on the order of 10\% of the poloidal component ($b_0\leq0.1$), that is oriented SE--NW (position angle $\varphi=-{44^\circ}_{-4^\circ}^{+6^\circ}$) and has an inclination $i=-{45^\circ}^{+3^\circ}_{-4^\circ}$, for a mass-to-flux ratio $\lambda$=2.66. The magnetic field is oriented perpendicular to the NE--SW velocity gradient detected in this core on scales from $\sim$10$^3$ to 10$^4$\,au consistent with our previous hypothesis that such a velocity gradient is due to rotation of the core. 

The almost poloidal geometry of the magnetic field in G31 is only perturbed by the collapse of both the Main core and the NE core, and by the molecular outflows detected in the core.  Notwithstanding the rotation of the core, this appears to have little effect on the magnetic field, as suggested by the fact that the best fit model includes only a very small toroidal component.

The strength of the magnetic field in the central region of the core has been estimated using the Davis-Chandrasekhar-Fermi method and is in the range $\sim$8\,mG to 13\,mG. This strength implies that the mass-to-flux ratio in this region is only slightly supercritical ($\lambda$=1.4--2.2). 
In fact, despite the magnetic field being important in G31, it is not sufficient to prevent fragmentation and collapse of the core, as demonstrated by the presence of (at least) four sources embedded in the Main core. The turbulent-to-magnetic energy ratio suggests that the two are comparable.

\begin{acknowledgements}  

We thank the anonymous referee and Philip Myers for their useful comments that have improved the manuscript. 
This paper makes use of the ALMA data ADS/JAO.ALMA\#2015.1.00072.S. ALMA is a partnership of ESO (representing its member states), NSF (USA) and NINS (Japan), together with NRC (Canada) and NSC and ASIAA (Taiwan) and KASI (Republic of Korea), in cooperation with the Republic of Chile. The Joint ALMA Observatory is operated by ESO, AUI/NRAO and NAOJ. MP acknowledges funding from the European Unions Horizon 2020 research and innovation programme under the Marie Skłodowska-Curie grant agreement No 664931. JMG acknowledges support from MICINN (Spain) AYA2017-84390-C2-2-R grant. GA and MO acknowledge financial support from the State Agency for Research of the Spanish MCIU through the AYA2017-84390-C2-1-R grant (co-funded by FEDER) and through the ``Center of Excellence Severo Ochoa'' award for the Instituto de Astrof\'{\i}sica de Andaluc\'{\i}a (SEV-2017-0709). 

 \end{acknowledgements}

\appendix

\section{Error on the observed polarization angles}\label{app:dpacontours}
\begin{figure}
\centerline{\includegraphics[angle=0,width=8.5cm,angle=0]{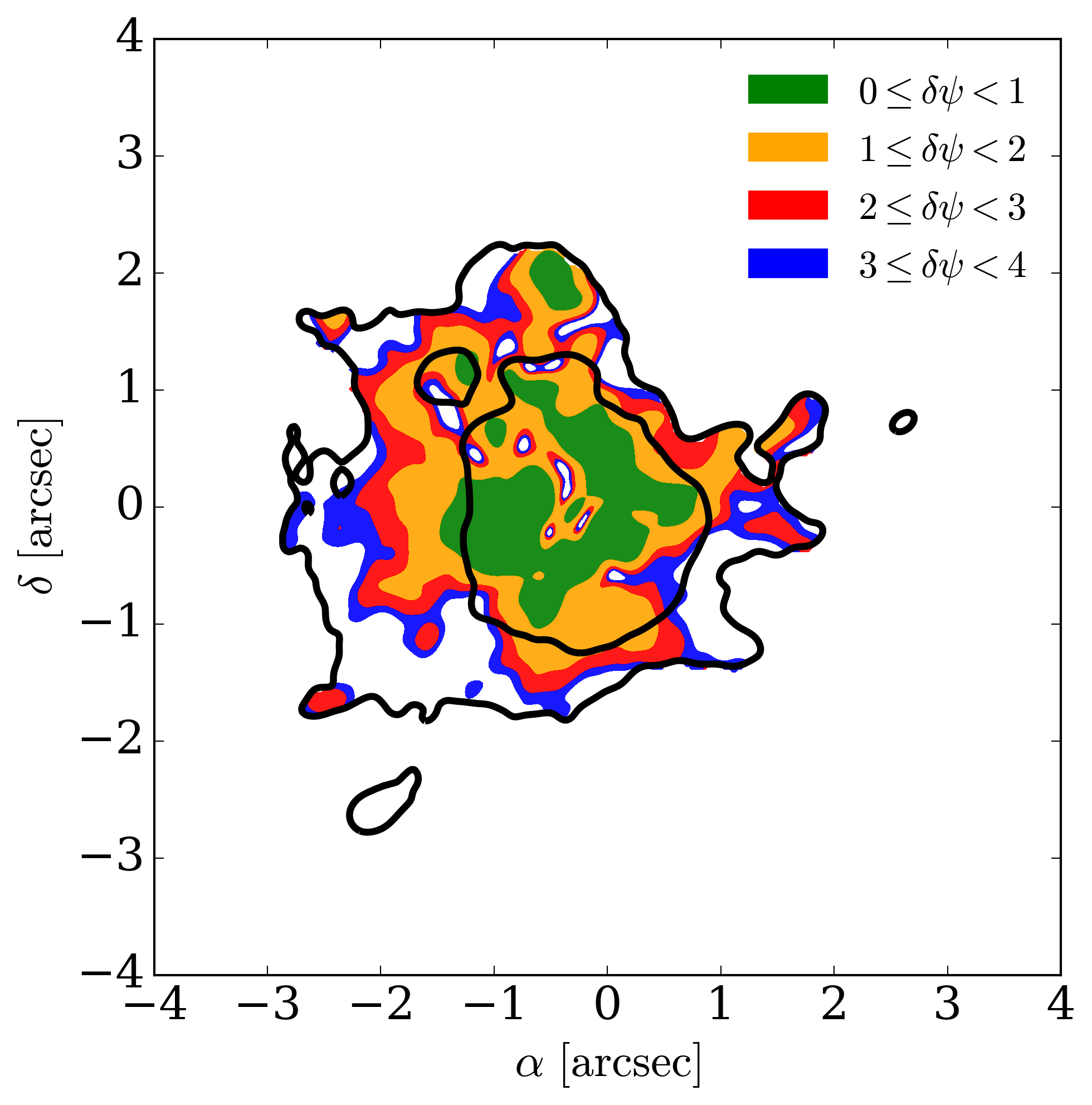}}
\caption{Map of the errors on the observed polarization angles according to the color-coding
shown in the legend. {\it Black contours} show the $5\sigma$ and $15\sigma$ levels of the 1.3~mm dust
continuum emission. {\it White regions} inside the $5\sigma$ contour refer to $\delta\psi_{\rm obs}\ge 4^\circ$.}
\label{dpacontours}
\end{figure}

The error on the polarization angle is computed by propagating the error from the definition
of $\psi$ (see Sect.~\ref{sect:obs}) and it is equal to
$\delta\psi_{\rm obs}=0.5\,\sigma_{\rm QU}/\sqrt{Q^2+U^2}$, 
where $\sigma_{\rm QU}=22~\mu$Jy is the noise on the observed Stokes Q and U. This expression of uncertainty in the polarization angle is valid for high ($>5$) signal-to-noise ratio (e.g., Vaillancourt~\cite{vaillancourt06}), which is the case of our data. As seen in Fig.~\ref{fig-pol} ({\it top panel}), the linearly polarized intensity $P$ of the G31 core is $>5\sigma_{\rm QU}$. Figure~\ref{dpacontours} shows that $\delta\psi_{\rm obs}$ is lower than 
about $4^\circ$ inside the 
$5\sigma$ contour of the 1.3~mm dust continuum emission and, on average, is lower than $2^\circ$
inside the $15\sigma$ contour encompassing the Main core.

\section{Chi-squared test for different values of the toroidal component}\label{app:chivsi266}
\begin{figure}
\centerline{\includegraphics[angle=0,width=8.5cm,angle=0]{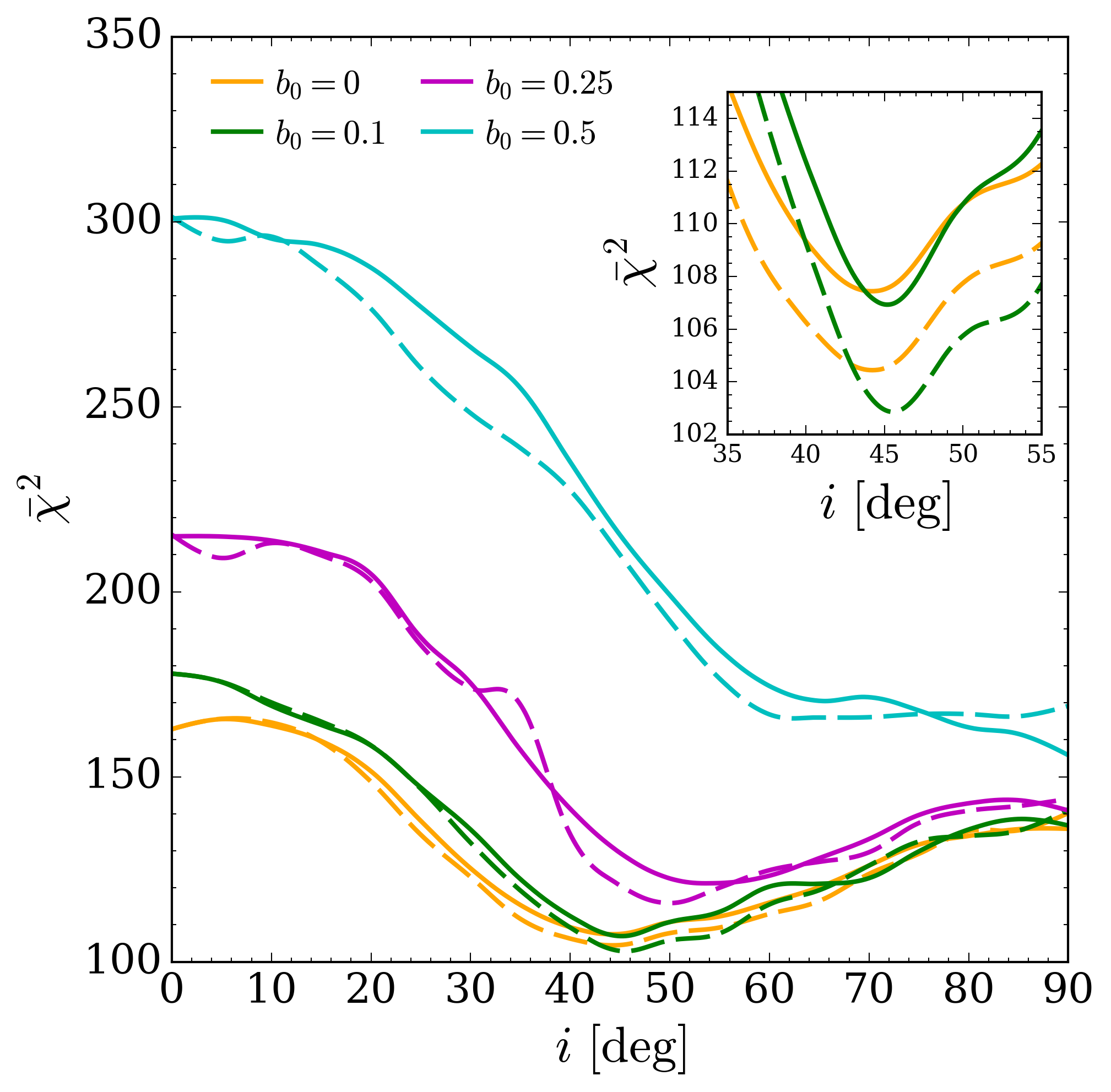}}
\caption{Minimum reduced chi squared versus the inclination, $i$, of the model for $\lambda=2.66$ and 
increasing toroidal component. 
{\it Solid} ({\it dashed}\/) lines show $\bar\chi^2$ for positive (negative) values of $i$.
For better comparison,
negative inclinations are shown in absolute value.
The inset shows a zoom around the minimum of $\bar\chi^2$ for
the cases $b_0=0$ and 0.1.}
\label{chi2vsincl_lambda266}
\end{figure}

Figure~\ref{chi2vsincl_lambda266} shows the values of $\bar\chi^2$ as
a function of the inclination for different values of $b_0$. This
test allowed us to discard the cases with $b_0>0.1$.

\end{document}